\documentclass{aa}
\usepackage[latin1]{inputenc}
\usepackage{graphicx}
\usepackage{amssymb}
\usepackage{mathrsfs}
\usepackage{natbib}
\bibpunct{(}{)}{;}{a}{}{,}

\begin{document}

\title{Rotational velocities of A-type stars \thanks{Based on observations made at the European Southern Observatory (ESO), La Silla, Chile, in the framework of the Key Programme 5-004-43K}\fnmsep\thanks{Table 4 is only available in electronic form at the CDS via anonymous ftp to cdsarc.u-strasbg.fr (130.79.125.5) or via http://cdsweb.u-strasbg.fr/Abstract.html}}
\subtitle{I. Measurement of $v\sin i$ in the southern hemisphere}

\author{F. Royer\inst{1,2}
   \and M. Gerbaldi\inst{3,4} 
   \and R. Faraggiana\inst{5}
   \and A.E. G\'omez\inst{2}}

\offprints{Frédéric Royer}

\institute{Observatoire de Gen\`eve, 51 chemin des Maillettes, CH-1290 Sauverny, Switzerland
      \and DASGAL/CNRS UMR 8633, Observatoire de Paris, 5 place Janssen, F-92195 Meudon cedex, France
      \and CNRS, Institut d'Astrophysique de Paris, 98 bis boulevard Arago, F-75014 Paris, France
      \and Universit\'e de Paris-Sud XI, F-91405 Orsay cedex, France
      \and Dipartimento di Astronomia, Universit\`a degli Studi di Trieste, via Tiepolo 11, I-34131 Trieste, Italy}

\date{Received date / Accepted date}

\titlerunning{Rotational velocities of A-type stars. I.}

%%%%%%%%%%%%%%%%%%%%%%%%%%%%%%%%
%%%%%%%%%%( ABSTRACT )%%%%%%%%%%
%%%%%%%%%%%%%%%%%%%%%%%%%%%%%%%%
\abstract{Within the scope of a Key Programme determining fundamental
  parameters of stars observed by HIPPARCOS, spectra of 525 B8 to
  F2-type stars brighter than $V=8$ have been collected at
  ESO. Fourier transforms of several line profiles in the range
  4200--4500\,\AA\ are used to derive \ensuremath{v\sin i}\ from the frequency of
  the first zero. Statistical analysis of the sample indicates that
  measurement error is a function of \ensuremath{v\sin i}\ and this relative error of the rotational velocity is found to be about 6\,\% on average.\\
The results obtained are compared with data from the literature. There
is a systematic shift from standard values from \citet{Slk_75}, which
are 10 to 12\,\% lower than our findings. Comparisons with other
independent \ensuremath{v\sin i}\ values tend to prove that those from Slettebak et
al. are underestimated. This effect is attributed to the presence of
binaries in the standard sample of Slettebak et al., and
to the model atmosphere they used.  
\keywords{techniques: spectroscopic -- stars: early-type; rotation}}
%%%%%%%%%%%%%%%%%%%%%%%%%%%%%%%%
 
\maketitle

%%%%%%%%%%%%%%%%%%%%%%%%%%%%%%%%
%%%%%%%%( INTRODUCTION )%%%%%%%%
%%%%%%%%%%%%%%%%%%%%%%%%%%%%%%%%
\section{Introduction}

Since work began on the subject \citep{SteEly31}, it has been observed that
stellar rotation rate is directly linked to the spectral type, and
A-type stars are known to be mean high rotators. 

The Doppler effect allows measurement of the broadening parameter \ensuremath{v\sin i}, the projection of the equatorial velocity $v$ along the line of sight. From a statistically significant sample of measured \ensuremath{v\sin i}, it is possible to derive the distribution of $v$ assuming that the rotation axes are randomly distributed and the sample is not biased.

Projected rotational velocities can be derived in many ways. Although large surveys of \ensuremath{v\sin i}\ already exist, great care must be
taken when combining their data, as various calibrations were used. 

The most accurate method of computing \ensuremath{v\sin i}\ would be the time-consuming computation of line profiles, starting from a model
atmosphere (with the introduction of other broadening mechanisms), and
their comparison with the observed lines \citep[see][ for their study
of Sirius]{Drs_90}. Such high precision is not justified, however, in a statistical study of high rotators like the non-peculiar A-type stars where other mechanisms (macroturbulence, instrumental) are negligible compared to rotation.

Line widths appear to be the natural indicator for measuring stellar
rotation, and most \ensuremath{v\sin i}\ are derived in this way, as a function of
the full-width at half-maximum (FWHM). The largest catalogue of
\ensuremath{v\sin i}\ is by \citet{UeiFua82}. It is an extremely heterogeneous
compilation of observational data mainly based on the old Slettebak
system \citep{Slk49,Slk54,Slk55,Slk56,SlkHod55}. Several years ago,
\citet{AbtMol95} measured \ensuremath{v\sin i}\ for 1700  A-type stars in the
northern hemisphere, calibrated with the new system from \citet[
hereafter SCBWP]{Slk_75}. More recently, \citet{WofSin97} measured the
\ensuremath{v\sin i}\ of 250 stars, most of which were cooler than those in our
sample, by cross-correlation with the spectra of standard stars of similar temperature. 
They found a small systematic difference with Abt \& Morrell's results 
(the former are larger by $\approx 5$~\%), and with those of \citet{DarFar72} (smaller by
8~\%). This can be explained by the difference between the ``old'' and ``new'' Slettebak systems. 
\citet{BrnVen97} derived \ensuremath{v\sin i}\ for a sample of early-type stars in
Sco~OB2 association from spectra taken with the same instrument we
used. They adopted three different techniques according to the
expected \ensuremath{v\sin i}\ values, which they show to be generally consistent
with each other. The \ensuremath{v\sin i}\ values so obtained correspond to those
defining the SCBWP scale, except for stars with \ensuremath{v\sin i}\ below 60\,\ensuremath{\mathrm{km}\,\mathrm{s}^{-1}}, for which the SCBWP values are systematically lower. 

The use of the Fourier technique in the determination of \ensuremath{v\sin i}\
remains occasional, mainly because using a calibration FWHM-\ensuremath{v\sin i}\ is
much easier and fitting theoretical profiles to observed ones in
wavelength space allows one to derive more parameters than simply the rotational broadening. Nevertheless, Fourier techniques are a valuable tool for investigating stellar rotation, as described by \citet{SmhGry76}.
\citet{Gry80b} compared the \ensuremath{v\sin i}\ obtained from Fourier transform of the \ion{Mg}{ii} 4481 line profile with the \ensuremath{v\sin i}\ values from Uesugi \& Fukuda and SCBWP and found a reasonable agreement (deviations of $\pm 10$\,\% with SCBWP), but his sample is quite small.

Suspecting that the small differences found with respect to standard 
values could be due to an underestimation in the SCBWP calibration of
the \ensuremath{v\sin i}\ values, we decided to undertake a measure of \ensuremath{v\sin i}\
independent of any pre-existing calibration. We adopted the method described in \citet{Raa_89}.

\paragraph{}The largest scatter in the average \ensuremath{v\sin i}\ distribution is
found for late B and early A stars \citep[ Fig.~17.16 p.~386]{Gry92},
and we want to test whether this is due only to errors in  measurement or if it is related to some physical effect.
\citet{BrnVen97}, in their study of the Sco~OB2 association, found
that B7-B9 stars of the Upper~Scorpius subgroup rotate faster than the
B0-B6 stars. This result corresponds to Gray's result, suggesting that
the apparent scatter may disguise a physical effect.  This effect has already been detected by \citet{Mos83}.

The possibility of a change on average \ensuremath{v\sin i}\ with evolution from zero-age to terminal-age main sequence has been studied for several decades, and the absence of any evolutionary effect for stars with a mass higher than $1.6\,\mathscr{M}_\odot$ is confirmed by the recent study of \citet{WofSin97}.
The fact that the colors of stars are affected by rotation was
observed for the first time by Brown \& Verschueren, but only for
stars belonging to young groups, not field stars. They conclude,
moreover, that the determination of ages and mass distributions is not affected by rotation.

As a matter of fact, the effect of rotation on stellar parameters is
also known: a rapidly rotating star simulates a star with lower \ensuremath{T_\mathrm{eff}}\
and $\log g$. However, in this case, all quantities (line strength,
photometric colors, for example) change in the same way so that the effect is practically undetectable \citep[this point was already discussed by][ p.~159]{Wof83}, especially when field stars are studied.

In this paper, newly determined \ensuremath{v\sin i}\ data, obtained with Fourier transforms, for 525 southern early-type stars are presented. The observations and the sample are described in Sect.~2. In Sect.~3 the technique used to derive \ensuremath{v\sin i}\ from the spectra is detailed and discussed. In Sect.~4 the results are presented and compared to data from the literature. In Sect.~5 our conclusions are summarized. 
This paper is the first of a series pertaining to rotational velocities of A-type stars; data collected in the northern hemisphere and measured \ensuremath{v\sin i}\ will be presented in a forthcoming paper.

%%%%%%%%%%%%%%%%%%%%%%%%%%%%%%%%
%%%%%%%%(    SAMPLE    )%%%%%%%%
%%%%%%%%%%%%%%%%%%%%%%%%%%%%%%%%
\section{Observational data}
\begin{figure}[!htp]
        \centering
\resizebox{\hsize}{!}{\includegraphics{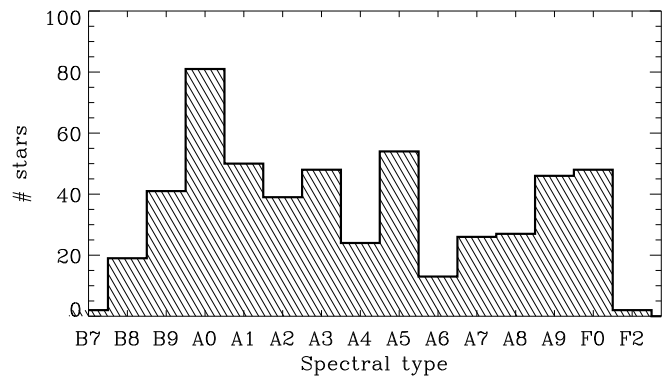}}  
        \caption{Distribution of the spectral type for the 525 programme stars.}
        \label{histST}
\end{figure}

The spectra were obtained with the ECHELEC spectrograph, mounted at
the Coud\'e focus of the ESO 1.52\,m telescope at La Silla. They were
collected from June 1989 to January 1995 in the framework of an ESO
Key Programme aimed at the determination of fundamental parameters of
early-type stars observed by HIPPARCOS \citep{GeiMar_89}, nearer than
100~pc. In total, 871 spectra were collected for 525 stars whose
spectral types range from B8 to F2 (Fig.~\ref{histST}). Most of these
stars belong to the main sequence (half of the sample are in luminosity class V, and a fifth is classified IV or IV-V).
These stars are all brighter than the $V$ magnitude 8.

It is worth noticing that the spectra which are the subject of the present paper
were also studied by \citet{Grr_99a} to derive radial velocities, and that
the 71 A0 dwarf stars observed were investigated by
\citet{Gei_99}. Basically, this sample includes objects with no
radial velocity or only for one epoch. Some stars with no \ensuremath{v\sin i}\
determination were added from the Bright Star Catalogue \citep{Bsc1}.
The observational programme is more detailed by \citet{Grr_99a}.

%%_________________________________________________
%%___________________________________________________
%%_________________________________________________
\begin{figure*}[!htp]
        \centering
\resizebox{\hsize}{!}{\includegraphics{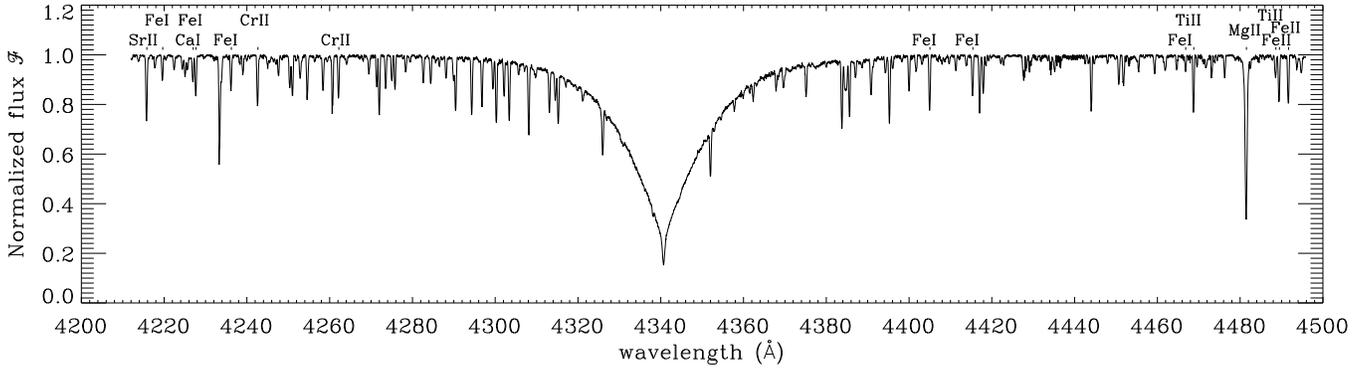}}  
        \caption{Normalized spectrum of Sirius, covering the 4210--4500 \AA\ range, around H$\gamma$. The 15 selected lines (listed in Table~\ref{raies_vsini}) are indicated}
        \label{spectre}
\end{figure*}
%%___________________________________________________
 The observations were made in the spectral range 4210--4500\,\AA\ (Fig.~\ref{spectre}).
The linear dispersion is about 3.1\,\AA\,$\mathrm{mm}^{-1}$, the slit width of
$320\,\mu$m corresponds to $1\farcs 52$ on the sky, and the power of resolution is about 28\,000.

The complete reduction of the spectra using MIDAS\footnote{MIDAS is being developed and maintained by ESO} package, from CCD frame to linear spectrum, is detailed by \citet{BueGei90,BueGei92}.

%%%%%%%%%%%%%%%%%%%%%%%%%%%%%%%%
%%%%%%%%(    METHOD    )%%%%%%%%
%%%%%%%%%%%%%%%%%%%%%%%%%%%%%%%%
\section{Measurement of the rotational velocity}

%===================
\subsection{Method}
%===================

\begin{figure}[!htp]
        \centering
\resizebox{\hsize}{!}{\includegraphics{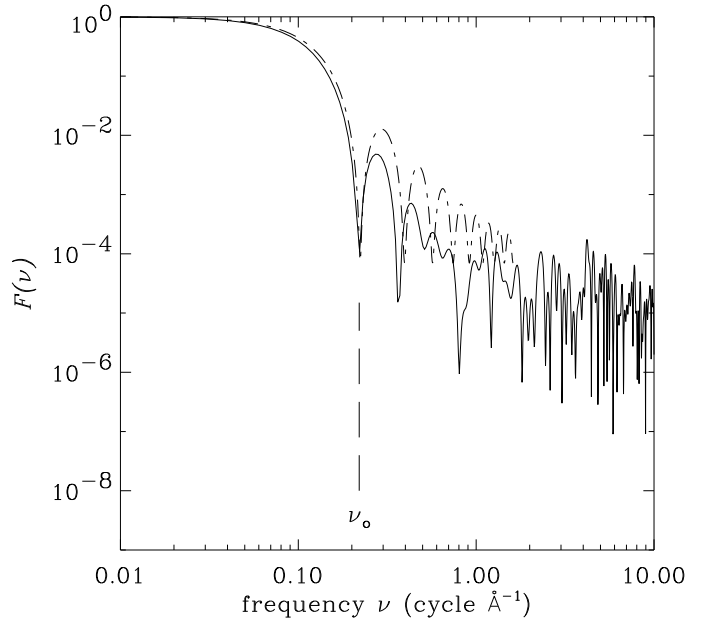}}  
        \caption{Profile of the Fourier transform of the \ion{Mg}{ii} 4481\,\AA\ line (solid line) for the star HIP~95965 and theoretical rotational profile (dashed line) with $\ensuremath{v\sin i}=200$\,\ensuremath{\mathrm{km}\,\mathrm{s}^{-1}}.}
        \label{fft_raie}
\end{figure}

As pointed out by \citet{GryGan87}, there is no ``standard'' technique for measuring projected rotational velocity.
The first application of Fourier analysis in the determination of stellar rotational velocities was undertaken by \citet{Cal33}.
\citet{Gry92} uses the whole profile of Fourier transform of spectral lines to derive the \ensuremath{v\sin i}, instead of only the zeroes as suggested by Carroll. 
The \ensuremath{v\sin i}\ measurement method we adopted is based on the position of
the first zero of the Fourier transform (FT) of the line profiles
\citep{Cal33}. The shape of the first lobe of the FT allows us to
better and more easily identify rotation as the main broadening agent of a line compared to the line profile in the wavelength domain. FT of the spectral line is computed using a Fast Fourier Transform algorithm. The \ensuremath{v\sin i}\ value is derived from the position of the first zero of the FT of the observed line using a theoretical rotation profile for a line at 4350\,\AA\ and \ensuremath{v\sin i}\ equal to 1\,\ensuremath{\mathrm{km}\,\mathrm{s}^{-1}}\ \citep{Raa_89}. The whole profile in the Fourier domain is then compared with a theoretical rotational profile for the corresponding velocity to check if the first lobes correspond (Fig.~\ref{fft_raie}). 
%%_________________________________________________
%%___________________________________________________

If $\nu_\mathrm{o}$ is the position of the first zero of the line profile (at $\lambda_\mathrm{o}$) in the Fourier space, the projected rotational velocity is derived as follows:
\begin{equation}
\label{Eq.vsini}
\ensuremath{v\sin i} = {4350\over\lambda_\mathrm{o}}\,{\nu_\mathrm{T}\over\nu_\mathrm{o}},
\end{equation} where 4350\,\AA\ and ${\nu_\mathrm{T}}$ respectively stand for the wavelength and the first zero of the theoretical profile.

It should be noted that we did not take into account the gravity
darkening, effect that can play a role in rapidly rotating stars when
velocity is close to break-up, as this is not relevant for most of our targets.

%=========================
\subsection{Continuum tracing}
%=========================

Determination of the projected rotational velocity requires normalized spectra.  

As far as the continuum is concerned, it has been determined visually, passing through noise fluctuations. The MIDAS procedure for continuum determination of 1D-spectra has been used, fitting a spline over the points chosen in the graphs.
 
Uncertainty related to this determination rises because the continuum observed on the spectrum is a pseudo-continuum. Actually, the true continuum is, in this spectral domain, not really reached for this type of stars. 
In order to quantify this effect, a grid of synthetic spectra of different effective temperatures (10\,000, 9200, 8500 and 7500\,K) and different rotational broadenings has been computed from Kurucz' model atmosphere \citep{Kuz93}, and Table~\ref{continuum} lists the differences between the true continuum and the pseudo-continuum represented as the highest points in the spectra.
%%______________________________________________________________
\begin{table*}[!htp]
\begin{center}
\caption{Differences between the true continuum and the highest points in different spectral bands for the set of synthetic spectra. }
\label{continuum}
\setlength\tabcolsep{3pt}
\begin{tabular}{r@{\hspace*{0.5mm}}rcccc|c|ccccc}
\hline
\multicolumn{2}{c}{\ensuremath{T_\mathrm{eff}}, \ensuremath{v\sin i}} & \multicolumn{10}{c}{spectral interval (\AA)} \cr
\multicolumn{2}{c}{(K, \ensuremath{\mathrm{km}\,\mathrm{s}^{-1}})}    & 4200--4220 & 4220--4240 & 4240--4260 & 4260--4280& H$\gamma$ & 4400--4420 & 4420--4440 & 4440--4460 & 4460--4480 & 4480--4500  \cr
\hline
10\,000, & 10 & 0.0029 & 0.0028 & 0.0035 & 0.0053 &                          & 0.0042 & 0.0023 & 0.0016 & 0.0012 & 0.0006 \cr
10\,000, & 50 & 0.0031 & 0.0044 & 0.0043 & 0.0073 &                          & 0.0057 & 0.0032 & 0.0023 & 0.0020 & 0.0011 \cr
10\,000, &100 & 0.0041 & 0.0059 & 0.0067 & 0.0077 &                          & 0.0063 & 0.0034 & 0.0040 & 0.0034 & 0.0022 \cr

   9200, & 10 & 0.0049 & 0.0049 & 0.0064 & 0.0090 &                          & 0.0068 & 0.0038 & 0.0027 & 0.0021 & 0.0012 \cr
   9200, & 50 & 0.0058 & 0.0071 & 0.0077 & 0.0123 &                          & 0.0085 & 0.0057 & 0.0046 & 0.0027 & 0.0023 \cr
   9200, &100 & 0.0076 & 0.0097 & 0.0110 & 0.0140 &                          & 0.0093 & 0.0063 & 0.0070 & 0.0050 & 0.0054 \cr

   8500, & 10 & 0.0073 & 0.0073 & 0.0100 & 0.0143 &                          & 0.0105 & 0.0059 & 0.0047 & 0.0032 & 0.0020 \cr
   8500, & 50 & 0.0099 & 0.0109 & 0.0133 & 0.0202 &                          & 0.0151 & 0.0105 & 0.0069 & 0.0044 & 0.0047 \cr
   8500, &100 & 0.0142 & 0.0152 & 0.0190 & 0.0262 &                          & 0.0156 & 0.0127 & 0.0140 & 0.0069 & 0.0114 \cr

   7500, & 10 & 0.0037 & 0.0037 & 0.0051 & 0.0069 &                          & 0.0055 & 0.0027 & 0.0023 & 0.0014 & 0.0009 \cr
   7500, & 50 & 0.0083 & 0.0101 & 0.0166 & 0.0225 &                          & 0.0132 & 0.0118 & 0.0064 & 0.0032 & 0.0068 \cr
   7500, &100 & 0.0212 & 0.0189 & 0.0257 & 0.0381 &                          & 0.0182 & 0.0191 & 0.0255 & 0.0064 & 0.0245 \cr
\hline
\end{tabular}
\end{center}
\end{table*}
%%______________________________________________________________
It illustrates the contribution of the wings of H$\gamma$, as the hydrogen lines reach their maximum strength in the early A-type stars, and in addition, the general strength of the metallic-line spectrum which grows with decreasing temperature.
In the best cases, i.e. earliest type and low broadening, differences are
about a few 0.1~\%. For cooler stars and higher rotators, they reach up to 3~\%.
The points selected to anchor the pseudo-continuum are selected as much as possible in the borders of the spectra, where the influence of the wings of H$\gamma$ is weaker.

Continuum is then tilted to origin and the spectral windows
corresponding to lines of interest are extracted from the spectrum in
order to compute their FT.

%=========================
\subsection{Set of lines}
%=========================

\subsubsection{{\em a priori} selection}

The essential step in this analysis is the search for suitable spectral lines to measure the \ensuremath{v\sin i}.
The lines which are candidates for use in the determination of rotation (Table~\ref{raies_vsini}) have been identified in the Sirius atlas \citep{Fud_92} and retained according to the following criteria:
\begin{itemize}
\item not blended in the Sirius spectrum,
\item far enough from the hydrogen line H$\gamma$ to maintain relatively good access to the continuum.
\end{itemize}
These are indicated in Fig.~\ref{spectre}.

%%______________________________________________________________
\begin{table}[!htp]
\begin{center}
\caption{List of the spectral lines used (when possible) for the \ensuremath{v\sin i}\ measurement.}
\label{raies_vsini}
\begin{tabular}{llll}
\hline
wavelength & element &\qquad wavelength & element\cr
 \multicolumn{1}{c}{(\AA)} &  &  \multicolumn{1}{c}{\qquad(\AA)} & \cr
\hline
4215.519 & \ion{Sr}{ii} &\qquad 4404.750 & \ion{Fe}{i}  \cr
4219.360 & \ion{Fe}{i}  &\qquad 4415.122 & \ion{Fe}{i}  \cr
4226.728 & \ion{Ca}{i}  &\qquad 4466.551 & \ion{Fe}{i}  \cr
4227.426 & \ion{Fe}{i}  &\qquad 4468.507 & \ion{Ti}{ii} \cr
4235.936 & \ion{Fe}{i}  &\qquad 4481$^{.126} _{.325}$ & \ion{Mg}{ii} \dag \cr
4242.364 & \ion{Cr}{ii} &\qquad 4488.331 & \ion{Ti}{ii} \cr
4261.913 & \ion{Cr}{ii} &\qquad 4489.183 & \ion{Fe}{ii} \cr
           &            &\qquad 4491.405 & \ion{Fe}{ii} \cr
\hline
\end{tabular}
\end{center}
\dag\ Wavelengths of both components are indicated for the magnesium doublet line.
\end{table}
%%__________________________________________________________

The lines selected in the Sirius spectrum are valid for early A-type
stars. When moving to stars cooler than about A3-type stars, the effects of
the increasing incidence of blends and the presence of stronger
metallic lines must be taken into account. The effects are: (1) an
increasing departure of the true continuum flux (to which the spectrum
must be normalized) from the curve that joins the highest points in
the observed spectrum, as mentioned in the previous subsection, and (2)
an increased incidence of blending that reduces the number of
lines suitable for \ensuremath{v\sin i}\ measurements. The former effect will be
estimated in Sect.~\ref{effectcontinuum}. The latter can be derived from the symmetry
of the spectral lines.
 
Considering a line, continuum tilted to zero, as a distribution, moments of $k$-th order can be defined as:
\begin{equation} 
\label{moment}
\forall k,\;m_k = {\displaystyle \sum_{i=1}^{L}\left[1-\mathscr{F}(\lambda_i)\right]  \left[\lambda_i-\lambda_\mathrm{c}\right]^k 
 \over \displaystyle \sum_{i=1}^{L} 1-\mathscr{F}(\lambda_i) },
\end{equation} for an absorption line centered at wavelength $\lambda_\mathrm{c}$ and spreading from $\lambda_1$ to $\lambda_L$, where $\mathscr{F}(\lambda_i)$ is the normalized flux corresponding to the wavelength $\lambda_i$. Ranges $[\lambda_1,\lambda_L]$ are centered around theoretical wavelengths from Table~\ref{raies_vsini} and the width of the window is taken to be 0.35, 0.90 and 1.80~\AA\ for rotational broadening 10, 50 and 100~\ensuremath{\mathrm{km}\,\mathrm{s}^{-1}}\ respectively (the width around the \ion{Mg}{ii} doublet is larger: 1.40, 2.0 and 2.3 \AA).
Skewness is then defined as
 
\begin{equation} 
\label{skew}
\gamma_1 = {m_3 \over (m_2)^{3/2}}.
\end{equation}

 Variations of skewness of a synthetic line profile with temperature
 and/or rotational broadening should be caused only by the presence of other spectral lines that distort the original profile. Table~\ref{skewness} gives skewness of the selected lines for the different synthetic spectra.
%%______________________________________________________________
\begin{table}[!htp]
\begin{center}
\caption{Variation of the skewness $\gamma_1$ (Eq.~\ref{skew}) of the lines with \ensuremath{T_\mathrm{eff}}\ and \ensuremath{v\sin i}\ in the synthetic spectra.}
\label{skewness}
\begin{tabular}{lrrrrr}
\hline
 & \multicolumn{1}{c}{\ensuremath{v\sin i}} & \multicolumn{4}{c}{\ensuremath{T_\mathrm{eff}}\ (K)} \cr
\cline{3-6}
line& \multicolumn{1}{c}{(\ensuremath{\mathrm{km}\,\mathrm{s}^{-1}})} &10\,000 &        9200   &         8500  &         7500 \cr
\hline   %-------|--------|-------------|---------------|---------------|---------------

\ion{Sr}{ii} 4216 &  10~~~~ & $ 0.07$	&	$ 0.10$	&	$ 0.18$	&	$ 0.22$ \\
                  &  50~~~~ & $ 0.04$	&	$ 0.08$	&	$ 0.18$	&	$ 0.36$ \\
                  & 100~~~~ & $ 0.05$	&	$ 0.10$	&	$ 0.17$	&       $ 0.24$ \\
\cline{2-6}
\ion{Fe}{i}  4219 &  10~~~~ & $ 0.03$	&	$ 0.03$	&	$ 0.04$	&	$ 0.05$ \\
                  &  50~~~~ & $ 0.03$	&	$ 0.08$	&	$ 0.18$	&	$ 0.32$ \\
                  & 100~~~~ & $-0.03$	&	$-0.05$	&	$-0.07$	&	$-0.03$ \\
\cline{2-6}
\ion{Ca}{i}  4227 &  10~~~~ & $-0.02$	&	$-0.07$	&	$-0.25$	&	$-0.63$ \\
                  &  50~~~~ & $-0.21$	&	$-0.28$	&	$-0.37$	&	$-0.49$ \\
                  & 100~~~~ & $-0.15$	&	$-0.22$	&	$-0.32$	&	$-0.40$ \\
\cline{2-6}
\ion{Fe}{i}  4227 &  10~~~~ & $-0.09$	&	$-0.13$	&	$-0.19$	&	$-0.28$ \\
                  &  50~~~~ & $-0.34$	&	$-0.48$	&	$-0.57$	&	$-0.69$ \\
                  & 100~~~~ & $-0.26$	&	$-0.36$	&	$-0.46$	&	$-0.55$ \\
\cline{2-6}
\ion{Fe}{i}  4236 &  10~~~~ & $-0.03$	&	$-0.06$	&	$-0.11$	&	$-0.24$ \\
                  &  50~~~~ & $-0.01$	&	$-0.03$	&	$-0.09$	&	$-0.24$ \\
                  & 100~~~~ & $-0.29$	&	$-0.30$	&	$-0.31$	&	$-0.29$ \\
\cline{2-6} 
\ion{Cr}{ii} 4242 &  10~~~~ & $ 0.01$	&	$-0.01$	&	$ 0.04$	&	$ 0.42$ \\
                  &  50~~~~ & $ 0.03$	&	$ 0.04$	&	$ 0.10$	&	$ 0.25$ \\
                  & 100~~~~ & $ 0.01$ 	&	$ 0.01$	&	$-0.03$	&	$-0.16$ \\
\cline{2-6} 
\ion{Cr}{ii} 4262 &  10~~~~ & $-0.07$	&	$-0.09$	&	$-0.13$	&	$-0.22$ \\
                  &  50~~~~ & $-0.19$	&	$-0.22$	&	$-0.24$	&	$-0.32$ \\
                  & 100~~~~ & $-0.27$	&	$-0.36$	&	$ 0.43$	&	$-0.50$ \\
\cline{2-6} 
\ion{Fe}{i}  4405 &  10~~~~ & $ 0.01$	&	$ 0.01$	&	$ 0.01$	&	$ 0.03$ \\
                  &  50~~~~ & $-0.00$	&	$-0.01$	&	$-0.01$	&	$-0.02$ \\
                  & 100~~~~ & $-0.03$  	&	$-0.03$	&	$-0.02$	&	$ 0.00$ \\
\cline{2-6}  
\ion{Fe}{i}  4415 &  10~~~~ & $ 0.10$	&	$ 0.21$	&	$ 0.35$	&	$ 0.42$ \\
                  &  50~~~~ & $ 0.10$	&	$ 0.17$	&	$ 0.23$	&	$ 0.22$ \\
                  & 100~~~~ & $ 0.30$	&	$ 0.33$	&	$ 0.33$	&	$ 0.28$ \\
\cline{2-6}
\ion{Fe}{i}  4467 &  10~~~~ & $ 0.00$	&	$-0.01$	&	$ 0.01$	&	$ 0.16$ \\
                  &  50~~~~ & $-0.02$	&	$-0.01$	&	$ 0.02$	&	$ 0.08$ \\
                  & 100~~~~ & $-0.13$ 	&	$-0.19$	&	$-0.26$	&	$-0.35$ \\
\cline{2-6}
\ion{Ti}{ii} 4468 &  10~~~~ & $-0.04$	&	$-0.05$	&	$-0.05$	&	$-0.05$ \\
                  &  50~~~~ & $ 0.13$	&	$ 0.27$	&	$ 0.42$	&	$ 0.52$ \\
                  & 100~~~~ & $ 0.10$	&	$ 0.13$	&	$ 0.20$	&	$ 0.28$ \\
\cline{2-6}
\ion{Mg}{ii} 4481 &  10~~~~ & $-0.06$	&	$ 0.04$	&	$ 0.45$	&	$ 0.97$ \\
                  &  50~~~~ & $-0.04$	&	$-0.00$	&	$ 0.11$	&	$ 0.27$ \\
                  & 100~~~~ & $-0.02$	&	$ 0.01$	&	$ 0.08$	&	$ 0.22$ \\
\cline{2-6}
\ion{Ti}{ii} 4488 &  10~~~~ & $-0.00$	&	$-0.03$	&	$-0.11$	&	$-0.31$ \\
                  &  50~~~~ & $-0.12$	&	$-0.16$	&	$-0.21$	&	$-0.26$ \\
                  & 100~~~~ & $-0.08$	&	$-0.10$	&	$-0.13$	&	$-0.18$ \\
\cline{2-6}
\ion{Fe}{ii} 4489 &  10~~~~ & $ 0.04$	&	$ 0.04$	&	$ 0.03$	&	$-0.02$ \\
                  &  50~~~~ & $-0.08$	&	$-0.12$	&	$-0.11$	&	$ 0.05$ \\
                  & 100~~~~ & $-0.20$	&	$-0.24$	&	$-0.28$	&	$-0.33$ \\
\cline{2-6}
\ion{Fe}{ii} 4491 &  10~~~~ & $ 0.01$	&	$ 0.01$	&	$ 0.02$	&	$ 0.03$ \\
                  &  50~~~~ & $-0.02$	&	$-0.05$	&	$-0.15$	&	$-0.31$ \\
                  & 100~~~~ & $-0.18$	&	$-0.23$	&	$-0.31$	&	$-0.41$ \\

\hline
\end{tabular}
\end{center}
\end{table}
%%__________________________________________________________

The most noticeable finding in this table is that $|\gamma_1|$ usually
increases with decreasing \ensuremath{T_\mathrm{eff}}\ and increasing \ensuremath{v\sin i}. This is a typical effect of blends. Nevertheless, high rotational broadening can lower the skewness of a blended line by making the blend smoother. 

Skewness $\gamma_1$ for the synthetic spectrum close to Sirius' parameters ($\ensuremath{T_\mathrm{eff}}=10\,000$\,K, $\ensuremath{v\sin i}=10$\,\ensuremath{\mathrm{km}\,\mathrm{s}^{-1}}) is contained between $-0.09$ and $+0.10$. The threshold, beyond which blends are regarded as affecting the profile significantly, is taken as equal to $0.15$. If $|\gamma_1|>0.15$ the line is not taken into account in the derivation of the \ensuremath{v\sin i}\ for a star with corresponding spectral type and rotational broadening. This threshold is a compromise between the unacceptable distortion of the line and the number of retained lines, and it ensures that the differences between centroid and theoretical wavelength of the lines have a standard deviation of about 0.02 \AA.

As can be expected, moving from B8 to F2-type stars increases the
blending of lines. Among the lines listed in Table~\ref{raies_vsini},
the strongest ones in Sirius spectrum (\ion{Sr}{ii} 4216, \ion{Fe}{i} 4219,
\ion{Cr}{ii} 4242, \ion{Fe}{i} 4405 and \ion{Mg}{ii} 4481) correspond
to those which remain less contaminated by the presence of other lines. Only \ion{Fe}{i} 4405 retains a symmetric profile not being heavily blended at the resolution of our spectra and thus measurable all across the grid of the synthetic spectra.

The \ion{Mg}{ii} doublet at 4481\,\AA\ is usually chosen to measure
the \ensuremath{v\sin i}: it is not very sensitive to stellar effective temperature
and gravity and its relative strength in late B through mid-A-type
star spectra makes it almost the only measurable line in this
%%_________________________________________________
\begin{figure}[htp]
        \centering
\resizebox{\hsize}{!}{\includegraphics{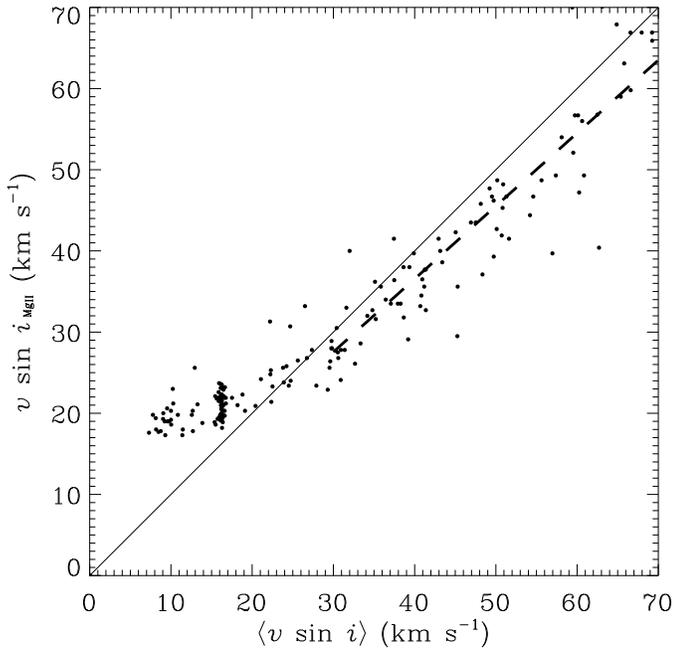}}  
        \caption{$\ensuremath{v\sin i}_\mathrm{\ion{Mg}{ii}}$ derived from the 4481
          \ion{Mg}{ii} line versus $\langle\ensuremath{v\sin i}\rangle$ derived from
          other metallic lines for early A-type stars. The solid line
          stands for the one-to-one relation. The dashed line is the
          least-squares linear fit for $\langle\ensuremath{v\sin i}\rangle>30$~\ensuremath{\mathrm{km}\,\mathrm{s}^{-1}}.}
        \label{vsini_MgII}
\end{figure}
%%___________________________________________________
spectral domain for high rotational broadening. However the separation
of 0.2\,\AA\ in the doublet leads to an overestimate of the \ensuremath{v\sin i}\ derived
from the \ion{Mg}{ii} line for low rotational
velocities. Figure~\ref{vsini_MgII} displays deviation between the
$\ensuremath{v\sin i}_{\mathrm{\ion{Mg}{ii}}}$ measured on the \ion{Mg}{ii} doublet
and the mean $\langle\ensuremath{v\sin i}\rangle$ derived from weaker metallic
lines, discarding automatically the \ion{Mg}{ii} line.
For low velocities, typically $\langle\ensuremath{v\sin i}\rangle\lesssim 25$\,\ensuremath{\mathrm{km}\,\mathrm{s}^{-1}},
the width of the doublet is not representative of the rotational
broadening but of the intrinsic separation between doublet
components. That is why $\ensuremath{v\sin i}_{\mathrm{\ion{Mg}{ii}}}$ is stagnant
at a plateau around 19\,\ensuremath{\mathrm{km}\,\mathrm{s}^{-1}}, and gives no indication of the true
rotational broadening. In order to take this effect into account,
the \ion{Mg}{ii} doublet is not used for \ensuremath{v\sin i}\ determination below
25\,\ensuremath{\mathrm{km}\,\mathrm{s}^{-1}}. 
For higher velocities, the \ensuremath{v\sin i}\ derived from weak lines are on the
average overestimated because they are prone to blending, whereas \ion{Mg}{ii} is
much more blend-free. On average, for $\langle\ensuremath{v\sin i}\rangle> 30$~\ensuremath{\mathrm{km}\,\mathrm{s}^{-1}}, 
the relation between $\langle\ensuremath{v\sin i}\rangle$ and
$\ensuremath{v\sin i}_{\mathrm{\ion{Mg}{ii}}}$ deviates from the one-to-one relation 
as shown by the dashed line in Fig.~\ref{vsini_MgII}, and a
least-squares linear fit gives the equation
\begin{equation} 
\label{mgii}
\ensuremath{v\sin i}_{\mathrm{\ion{Mg}{ii}}} = 0.9\,\langle\ensuremath{v\sin i}\rangle + 0.6,
\end{equation}
which suggests that blends can lead to a 10~\% overestimation of the \ensuremath{v\sin i}.

\subsubsection{{\em a posteriori} selection}

Among the list of candidate lines chosen according to the spectral type and rotational broadening of the star, some can be discarded on the basis of the spectrum quality itself. The main reason for discarding a line, first supposed to be reliable for \ensuremath{v\sin i}\ determination, lies in its profile in Fourier space. One retains the results given by lines whose profile correspond to a rotational profile.

\begin{figure}[!htp]
        \centering
\resizebox{\hsize}{!}{\includegraphics{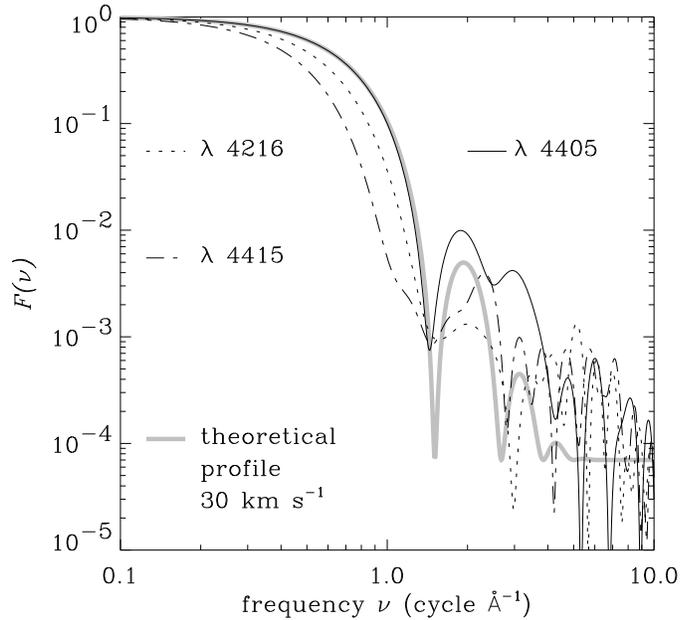}}  
        \caption{Example of line profiles in the Fourier space for HD~75063 (A1III type star) whose $\ensuremath{v\sin i}=30\,\ensuremath{\mathrm{km}\,\mathrm{s}^{-1}}$. The theoretical rotational profile (grey solid line), computed for the average \ensuremath{v\sin i}\ of the star, matches perfectly the FT of the \ion{Fe}{i} 4405 line (black solid line), whereas Fourier profiles of \ion{Sr}{ii} 4216 and \ion{Fe}{i} 4415 differ from a rotational shape. Among these three observed lines, only \ion{Fe}{i} 4405 is retained for \ensuremath{v\sin i}\ determination.}
        \label{fft_raie3}
\end{figure}

In logarithmic frequency space, such as in Figs.~\ref{fft_raie} and~\ref{fft_raie3}, the rotational profile has a unique shape, and the effect of \ensuremath{v\sin i}\ simply acts as a translation in frequency. Matching between the theoretical profile, shifted at the {\em ad hoc} velocity, and the observed profile, is used as confirmation of the value of the first zero as a \ensuremath{v\sin i}. 
%%_________________________________________________
%%___________________________________________________
This comparison, carried out visually, allows us to discard non suitable Fourier profiles as shown in Fig.~\ref{fft_raie3}. 

A discarded Fourier profile is sometimes associated with a distorted profile in wavelength space, but this is not always the case. For low rotational broadening, i.e. $\ensuremath{v\sin i}\lesssim 10$~\ensuremath{\mathrm{km}\,\mathrm{s}^{-1}}, the Fourier profile deviates from the theoretical rotational profile. This is due to the fact that rotation does not completely dominate the line profile and the underlying instrumental profile is no longer negligible. It may also occur that an SB2 system, where lines of both components are merged, appears as a single star, but the blend due to multiplicity makes the line profile diverge from a rotational profile.

\paragraph{}To conclude, the number of measurable lines among the 15 listed in Table~\ref{raies_vsini} also varies from one spectrum to another according to the rotational broadening and the signal-to-noise ratio and ranges from 1 to 15 lines. 
%%_________________________________________________
\begin{figure}[!htp]
        \centering
\resizebox{\hsize}{!}{\includegraphics{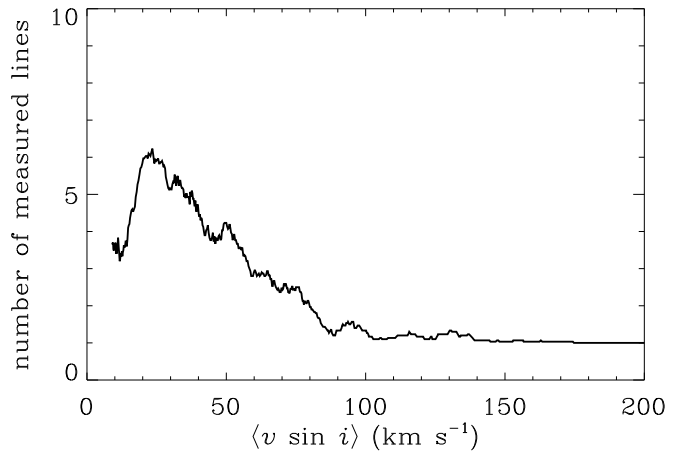}}  
        \caption{Average number of measured lines (running average over 30 points) is plotted as a function of the mean $\langle\ensuremath{v\sin i}\rangle$.}
        \label{line-vsini}
\end{figure}
%%___________________________________________________
As shown in Fig.~\ref{line-vsini}, the average number of measured lines 
decreases almost linearly with increasing \ensuremath{v\sin i}, because of blends, and reaches one (the 
\ion{Mg}{ii} 4481 doublet line) at $\ensuremath{v\sin i}\approx 100$~\ensuremath{\mathrm{km}\,\mathrm{s}^{-1}}. Below
about 25~\ensuremath{\mathrm{km}\,\mathrm{s}^{-1}}, the number of measured lines decreases with \ensuremath{v\sin i}\ for 
two reasons: first, \ion{Mg}{ii} line is not used due to its intrinsic 
width; and more lines are discarded because of their 
non-rotational Fourier profile, instrumental profile being less negligible.

%==============================================
\subsection{Systematic effect due to continuum}
%==============================================
\label{effectcontinuum}
The measured continuum differs from the true one, and the latter is generally underestimated due to the wings of H$\gamma$ and the blends of weak metallic lines. One expects a systematic effect of the pseudo-continuum on the \ensuremath{v\sin i}\ determination as the depth of a line appears lower, and so its FWHM.
 
We use the grid of synthetic spectra to derive rotational broadening from ``true normalized'' spectra (directly given by the models) and ``pseudo normalized'' spectra (normalized in the same way as the observed spectra). The difference of the two measurements is
\begin{equation}
\delta\ensuremath{v\sin i}= \ensuremath{v\sin i} - \ensuremath{v\sin i}',
\end{equation} where \ensuremath{v\sin i}\ is the rotational broadening derived using a pseudo-continuum and $\ensuremath{v\sin i}'$ using the true continuum.
%%_________________________________________________
\begin{figure}[!htp]
        \centering
\resizebox{\hsize}{!}{\includegraphics{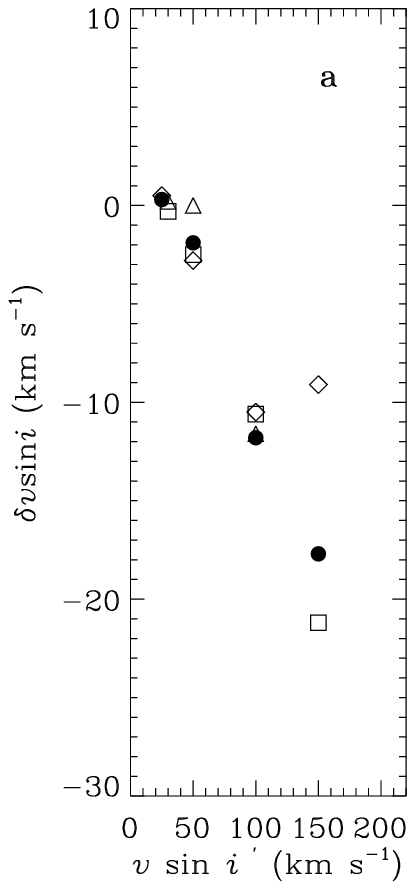}\includegraphics{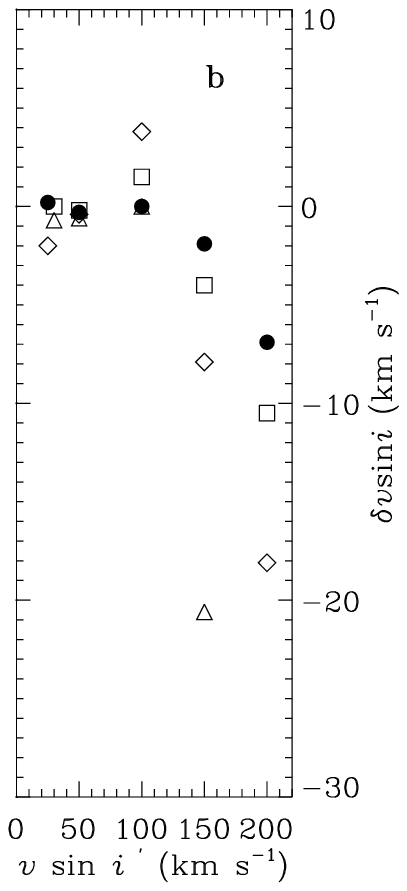}}  
        \caption{Systematic shift $\delta\ensuremath{v\sin i}$ as a function of the true rotational broadening $\ensuremath{v\sin i}'$ for:  {\bf a} \ion{Fe}{i} 4405 and {\bf b} \ion{Mg}{ii} 4481. The different symbols stand for the effective temperature of the synthetic spectra: fill circle: 10\,000~K; square: 9200~K; diamond: 8500~K; triangle: 7500~K.}
        \label{dvsinivsini}
\end{figure}
%%___________________________________________________
The systematic effect of the normalization induces an underestimation
of the \ensuremath{v\sin i}\ as shown in Fig.~\ref{dvsinivsini}. This shift depends
on the spectral line and its relative depth compared to the difference
between true and pseudo-continuum. For \ion{Fe}{i} 4405
(Fig.~\ref{dvsinivsini}.a), the shift is about $0.15\,(\ensuremath{v\sin i}'-30)$,
which leads to large differences. The effect is quite uniform in
temperature as reflected by the symmetric shape of the line all along
the spectral sequence (Table~\ref{skewness}). Nevertheless, at
$\ensuremath{v\sin i}'=150$~\ensuremath{\mathrm{km}\,\mathrm{s}^{-1}}, the scattering can be explained by the strength of
H$\gamma$ whose wing is not negligible compared to the flattened
profile of the line. For \ion{Mg}{ii} 4481 (Fig.~\ref{dvsinivsini}.b),
the main effect is not due to the Balmer line but to blends of metallic
lines. The shift remains small for early A-type stars (filled circles
and open squares):
$\lesssim 5$~\%, but increases with decreasing effective temperature
of the spectra, up to 10~\%. This is due to the fact that \ion{Mg}{ii} 
doublet is highly affected by blends for temperatures cooler than about
9000~K.

This estimation of the effect of the continuum is only carried out on
synthetic spectra because the way our observed spectra have been
normalized offers no way to recover the true continuum. The resulting
shift is given here for information only.

%======================
\subsection{Precision}
%======================
\label{precision}

Two types of uncertainties are present: those internal to the method and those related to the line profile. 

The internal error comes from the uncertainty in the real position of the first zero due to the sampling in the Fourier space. The Fourier transforms are computed over 1024 points equally spaced with the step $\Delta\nu$. This step is inversely proportional to the step in wavelength space $\Delta\lambda$, and the spectra are sampled with $\Delta\lambda=0.05$~\AA. The uncertainty of \ensuremath{v\sin i}\ due to the sampling is 
\begin{equation}
\label{samplingFFT}
\begin{array}{ccl}
\Delta\ensuremath{v\sin i} & \propto & (\ensuremath{v\sin i})^2\,\lambda_\mathrm{o}\,\Delta\nu\\
             & \propto & (\ensuremath{v\sin i})^2\,{\lambda_\mathrm{o}\over\Delta\lambda}\\
             & \approx & 4.\,10^{-4}\,(\ensuremath{v\sin i})^2.
\end{array}
\end{equation}

This dependence with \ensuremath{v\sin i}\ to the square makes the sampling step very small for low \ensuremath{v\sin i}\ and it reaches about 1\,\ensuremath{\mathrm{km}\,\mathrm{s}^{-1}}\ for $\ensuremath{v\sin i}=50$\,\ensuremath{\mathrm{km}\,\mathrm{s}^{-1}}.

The best way to estimate the precision of our measurements is to study the dispersion of the individual \ensuremath{v\sin i}. For each star, \ensuremath{v\sin i}\ is an average of the individual values derived from selected lines. 

\subsubsection{Effect of \ensuremath{v\sin i}}
 
\begin{figure}[!htp]
        \centering
\resizebox{\hsize}{!}{\includegraphics{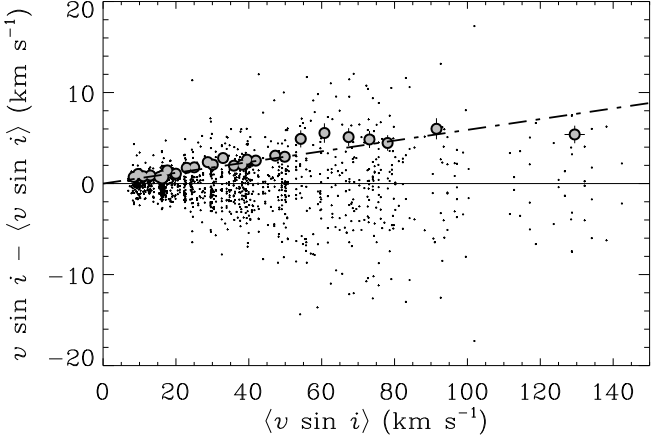}}  
        \caption{Differences between individual \ensuremath{v\sin i}\ and mean over
          a spectrum $\langle\ensuremath{v\sin i}\rangle$. Variation of the standard
          deviation associated with the measure as a function of
          $\langle\ensuremath{v\sin i}\rangle$ is shown by the open circles. A
          linear least squares fit on these points (dot-dashed line) gives a slope of 0.06.}
        \label{sigma-vsini1}
\end{figure}

The error associated with the \ensuremath{v\sin i}\ is expected to depend on \ensuremath{v\sin i},
because Doppler broadening makes the spectral lines shallow; that is,
it reduces the contrast line/continuum and increases the occurrence of
blends. Both effects tend to disrupt the selection of the lines as
well as the access to the continuum. Moreover, the stronger the
rotational broadening is, the fewer measurable lines there are. In Fig.~\ref{sigma-vsini1}, the differences between the individual \ensuremath{v\sin i}\ values from each measured line in each spectrum with the associated mean value for the spectrum are plotted as a function of $\langle\ensuremath{v\sin i}\rangle$. 
%%_________________________________________________
%%___________________________________________________
A robust estimate of the standard deviation is computed for each bin of 50 points; resulting points (open grey circles in Fig.~\ref{sigma-vsini1}) are adjusted with a linear least squares fit (dot-dashed line) giving:
\begin{equation}
\label{sigma}
\sigma_{\ensuremath{v\sin i}|\ensuremath{v\sin i}}=0.059{\scriptstyle\pm 0.003}\,\langle\ensuremath{v\sin i}\rangle. 
\end{equation} 
The fit is carried out using GaussFit \citep{Jes_98a,Jes_98b}, a
general program for the solution of least squares and robust
estimation problems. The formal error is then estimated as 6\,\% of
the \ensuremath{v\sin i}\ value.

\subsubsection{Effect of spectral type}

%%_________________________________________________
\begin{figure}[!htp]
        \centering
\resizebox{\hsize}{!}{\includegraphics{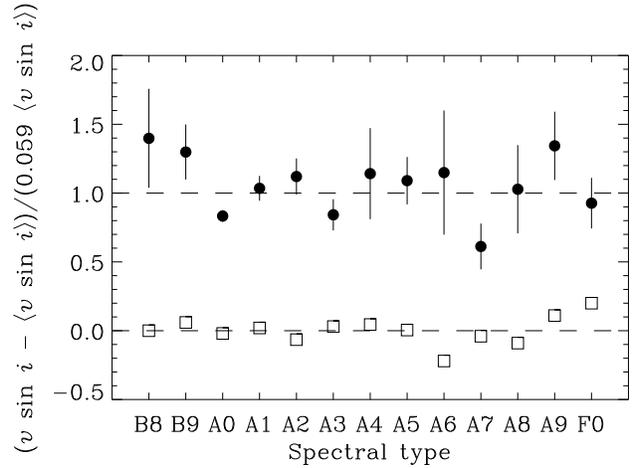}}  
        \caption{Mean of differences between individual \ensuremath{v\sin i}\ and
          average $\langle\ensuremath{v\sin i}\rangle$ over a spectrum, normalized
          by the formal error due to \ensuremath{v\sin i}, are indicated for each
          spectral type by the open squares. The standard deviations
          of these means for each spectral class are plotted as filled 
          circles, with their associated error bar.}
        \label{sigma-vsini3}
\end{figure}
%%___________________________________________________

Residual around this formal error can be expected to depend on the
effective temperature of the star. Figure~\ref{sigma-vsini3} displays
the variations of the residuals as a function of the spectral
type. Although contents of each bin of spectral type are not constant
all across the sample (the error bar is roughly proportional to the
logarithm of the inverse of the number of points), there does not seem to be
any trend, which suggests that our choice of lines according to the
spectral type eliminates any systematic effect due to the stellar temperature from the measurement
of the \ensuremath{v\sin i}.

\subsubsection{Effect of noise level}
Although noise is processed as a high frequency signal by Fourier technique and not supposed to act much upon \ensuremath{v\sin i}\ determination from  the first lobe of the FT, signal-to-noise ratio (SNR) may affect the measurement.
SNR affects the choice of the lines' limits in the spectrum as well as the computation of the lines' central wavelength.
%%_________________________________________________
\begin{figure}[!htp]
        \centering
\resizebox{\hsize}{!}{\includegraphics{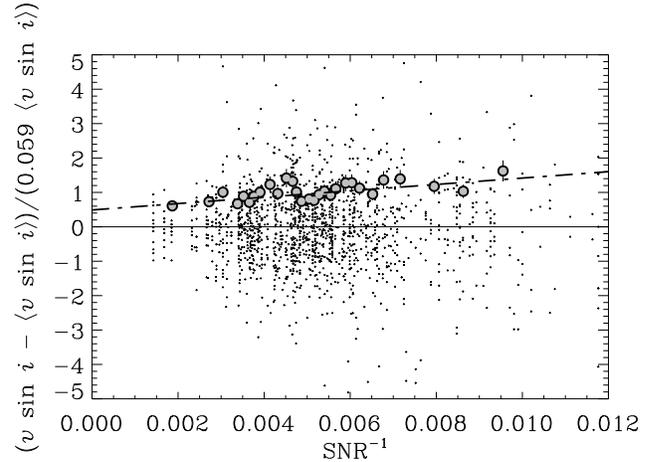}}  
        \caption{Differences between individual \ensuremath{v\sin i}\ and mean over
          a spectrum $\langle\ensuremath{v\sin i}\rangle$, normalized by the formal
          error due to \ensuremath{v\sin i}. Variation of the standard deviation
          associated to the measure with the noise level (SNR$^{-1}$)
          is shown by the open circles. A linear least squares fit on these points (dot-dashed line) gives a slope of $\sim 100$}
        \label{sigma-vsini2}
\end{figure}
%%___________________________________________________

The differences $\ensuremath{v\sin i} - \langle\ensuremath{v\sin i}\rangle$, normalized by the
formal error $0.059\,\langle\ensuremath{v\sin i}\rangle$, are plotted versus the
noise level (SNR$^{-1}$) in Fig.~\ref{sigma-vsini2} in order to
estimate the effect of SNR. Noise is derived for each spectrum using a
piecewise-linear high-pass filter in Fourier space with a transition
band chosen between $0.3$ and $0.4$ times the Nyquist frequency;
standard deviation of this high frequency signal is computed as the
noise level and then divided by the signal level. The trend in
Fig.~\ref{sigma-vsini2} is computed as for Fig.~\ref{sigma-vsini1},
using a robust estimation and GaussFit. 
The linear adjustment gives:
\begin{equation}
\label{sigma2}
\hat{\sigma}_{\ensuremath{v\sin i}|\mathrm{SNR}} = 93{\scriptstyle\pm 16}\,\mathrm{SNR}^{-1} + 0.5{\scriptstyle\pm 0.1}. 
\end{equation} 

The distribution of mean signal-to-noise ratios for our observations
peaks at $\mathrm{SNR} = 190$ with a standard deviation of $78$. This means that for most of the observations, SNR does not contribute much to the formal error on \ensuremath{v\sin i}\ ($\hat{\sigma}_{\ensuremath{v\sin i}|\mathrm{SNR}} \approx 1$).
Finally, the formal error associated with the \ensuremath{v\sin i}\ can be quantified as:
\begin{equation}
\label{sigma3}
\begin{array}{ccl}
\sigma_{\ensuremath{v\sin i}} & = & \sigma_{\ensuremath{v\sin i}|\ensuremath{v\sin i}}\,\hat{\sigma}_{\ensuremath{v\sin i}|\mathrm{SNR}} \\
                & = & \left(0.059\,\langle\ensuremath{v\sin i}\rangle\right)\,\left(93\,\mathrm{SNR}^{-1} + 0.5\right)\\
                &\propto & {\ensuremath{v\sin i}\over\mathrm{SNR}}.
\end{array} 
\end{equation} 
 
\subsubsection{Error distribution}

Distribution of observational errors, in the case of rotational
velocities, is of particular interest during a deconvolution process
in order to get rid of statistical errors in a significant sample.

\begin{figure}[!htp]
        \centering
\resizebox{\hsize}{!}{\includegraphics{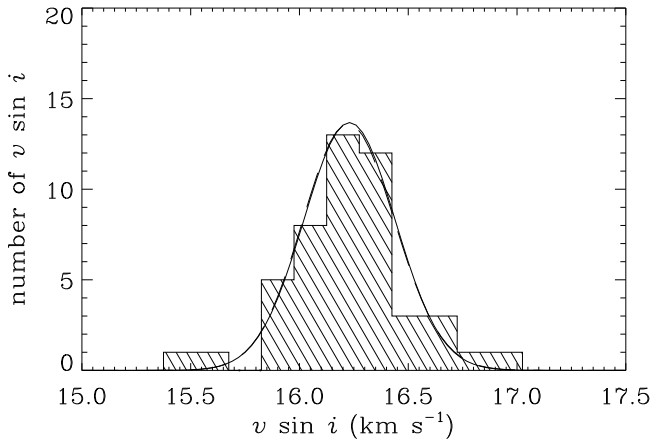}}  
        \caption{The \ensuremath{v\sin i}\ determinations for the 48 spectra of Sirius are distributed around the mean $16.22{\scriptstyle\pm 0.04}$\,\ensuremath{\mathrm{km}\,\mathrm{s}^{-1}}\ with a dispersion of $0.27{\scriptstyle\pm 0.03}$\,\ensuremath{\mathrm{km}\,\mathrm{s}^{-1}}. The optimal normal distribution $\mathscr{N}(16.23,0.21)$ that fits the histogram with a 96\,\% significance level is over-plotted. The optimal log-normal distribution merges together with the gaussian.}
        \label{histsirius}
\end{figure}

To have an idea of the shape of the error law associated with the \ensuremath{v\sin i}, it is necessary to have a great number of spectra for the same star.
\object{Sirius} has been observed on several occasions during the runs and its spectrum has been collected 48 times. Sirius spectra typically exhibit high signal-to-noise ratio (SNR\,$\gtrsim 250$). The 48 values derived from each set of lines, displayed in Fig.~\ref{histsirius}, give us an insight into the errors distribution. The mean \ensuremath{v\sin i}\ is $16.22{\scriptstyle\pm 0.04}$\,\ensuremath{\mathrm{km}\,\mathrm{s}^{-1}}\ and its associated standard deviation $0.27{\scriptstyle\pm 0.03}$\,\ensuremath{\mathrm{km}\,\mathrm{s}^{-1}}; data are approximatively distributed following a gaussian around the mean \ensuremath{v\sin i}.
%%_________________________________________________
%%___________________________________________________
A Kolmogorov-Smirnov test shows us that, with a 96\,\% significance level, this distribution is not different from a gaussian centered at 16.23 with a standard deviation equal to 0.21.
In the case of Sirius (low \ensuremath{v\sin i}) the error distribution corresponds to a normal distribution. We may expect a log-normal
distribution as the natural error law, considering that the error on \ensuremath{v\sin i}\ is multiplicative. But for low \ensuremath{v\sin i}, and low dispersion, log-normal and normal distributions do not significantly differ from each other.

 Moreover, for higher broadening, the impact of the sampling effect of
 the FT (Eq.~\ref{samplingFFT}) is foreseen, resulting in a distribution with a box-shaped profile. This effect becomes noticeable for $\ensuremath{v\sin i}\gtrsim 100$~\ensuremath{\mathrm{km}\,\mathrm{s}^{-1}}.

%%%%%%%%%%%%%%%%%%%%%%%%%%%%%%%%
%%%%%%%%(  VSINI DATA  )%%%%%%%%
%%%%%%%%%%%%%%%%%%%%%%%%%%%%%%%%

\section{Rotational velocities data}

%====================
\subsection{Results}
%====================

In all, projected rotational velocities were derived for 525 B8 to
F2-type stars. Among them, 286 have no rotational velocities either
in the compilation of \citet{UeiFua82} or in \citet{AbtMol95}.

The results of the \ensuremath{v\sin i}\ determinations are presented in
Table~\ref{results} which contains the following data: column (1)
gives the HD number, column (2) gives the HIP number, column (3)
displays the spectral type as given in the HIPPARCOS catalogue
\citep{Hip}, columns (4,5,6) give respectively the derived value of
\ensuremath{v\sin i}, the associated standard deviation and the corresponding number
of measured lines (uncertain \ensuremath{v\sin i}\ are indicated by a colon), column
(7) presents possible remarks about the spectra: SB2 (``SB'') and
shell (``SH'') natures are indicated for stars detailed in the
subsections which follow, as 
well as the reason why \ensuremath{v\sin i}\ is uncertain -- ``NO'' for no selected
lines, ``SS'' for variation from spectrum to spectrum and ``LL'' for
variation from line to line, as detailed in the Appendix~\ref{Notes0}.

\citet{Grr_99a} studied the same stars with the same spectra and derived radial velocities using cross-correlation techniques. On the basis of the shape of the cross-correlation function (CCF) they find that less than half of the sample has a symmetric and gaussian CCF and they classify stars with distorted CCF as, among other things, ``certain'' ``probable'' or ``suspected'' doubles.
%%______________________________________________________________
\begin{table}[!htp]
\centering
\caption{{\bf (extract)} Results of the \ensuremath{v\sin i}\ measurements. Only the 15 first stars are listed below. The whole table is available electronically.}
\label{results}
\begin{tabular}{rrlrrrl}
\hline
\multicolumn{1}{c}{HD} & \multicolumn{1}{c}{HIP} & Spect. type & \ensuremath{v\sin i}\ & $\sigma$ & \# & Remark\\
                       &                         &             &  \multicolumn{2}{r}{(\ensuremath{\mathrm{km}\,\mathrm{s}^{-1}})~}  \\
\hline

   256 &    602 & A2IV/V       & 243 &  9 &   2 & \\
   319 &    636 & A1V          &  61 &  5 &   8 & \\
   560 &    813 & B9V          & 249 & -- &   1 & \\
   565 &    798 & A6V          & 149 &  0 &   2 & \\
  1064 &   1191 & B9V          & 128 & -- &   1 & \\
  1685 &   1647 & B9V          & 236 & -- &   1 & \\
  1978 &   1900 & A0           & 110:& -- &   1 & NO\\
  2696 &   2381 & A3V          & 168 &  3 &   2 & \\
  3003 &   2578 & A0V          &  93 & -- &   1 & \\
  3652 &   3090 & F0           &  46 & -- &   1 & \\
  4065 &   3356 & B9.5V        &  35 &  4 &   4 & \\
  4125 &   3351 & A6V          &  90 & -- &   1 & \\
  4150 &   3405 & A0IV         & 124 &  0 &   2 & \\
  4338 &   3576 & F0IV         &  75 & -- &   1 & \\
  4772 &   3858 & A3IV         & 157 &  3 &   2 & \\

\hline
\end{tabular}
\end{table}
%%__________________________________________________________

Uncertainties in \ensuremath{v\sin i}\ are induced by peculiarities in the spectra
due for example to binarity or to the presence of a shell. The results for these objects are detailed below. These objects were either known as binaries or newly detected by \citet{Grr_99a}.

\subsubsection{Binary systems}
Spectra of double-lined spectroscopic binary systems (SB2) display lines of both components of the system. They are, by nature, more affected by blends and require much more attention than single stars in order to disentangle both spectra.

Moreover, the difference in radial velocity $\Delta V_\mathrm{r}$ has
to be large enough for the spectrum to show well separated
lines. Considering a gaussian line profile, 98\,\% of the distribution
is contained between $\pm 2.326\,\sigma$ ($\sigma$ being the standard
deviation of the Gaussian) which is nearly equal to
$\pm\mathrm{FWHM}$. It follows that a double line resulting from the contribution of the components of a binary system should be spaced of $|\Delta\lambda_\mathrm{A}-\Delta\lambda_\mathrm{B}| \gtrsim 2\,\mathrm{FWHM}$ (where $\Delta\lambda_\mathrm{A}$ and $\Delta\lambda_\mathrm{B}$ are the respective Doppler shifts) to overlap as little as possible and be measurable in terms of \ensuremath{v\sin i}\ determination. Taking the calibration relation from SCBWP as a rule of thumb ($\mathrm{FWHM}{\scriptstyle [\mathrm{\AA}]} \approx 0.025\,\ensuremath{v\sin i}{\scriptstyle [\ensuremath{\mathrm{km}\,\mathrm{s}^{-1}}]}$), the difference of radial velocity in an SB2 system should be higher than:
 \begin{equation}
\label{DeltaVr}
\Delta V_\mathrm{r} \gtrsim  {2\,c\,0.025\over \lambda}\,\ensuremath{v\sin i} \approx 3.4\,\ensuremath{v\sin i},
\end{equation} where $c$ is the velocity of light and $\lambda$ the
wavelength of the line ($\sim 4350$\,\AA). This threshold is a rough
estimate of whether \ensuremath{v\sin i}\ is measurable in the case of SB2.
On the other hand, the respective cores of the double line cease to be distinct when relative Doppler shift is lower than the FWHM, considering gaussian profiles\footnote{The sum of two identical Gaussians separated with $\Delta$ does not show splitted tops when 
\begin{eqnarray*}
\Delta & < & 2\,\sigma\,\sqrt{-2\,\log \left[ {1\over 3} {( 17+3\,\sqrt{33} ) }^{1/3} -{2\over 3\,{( 17+3\,\sqrt{33}) }^{1/3}}-{1\over 3}\right] } \\
       & < & 2.208\,\sigma \\
       & < & 0.94\,\mathrm{FWHM}.
\end{eqnarray*}}, i.e. $\Delta V_\mathrm{r}\lesssim 1.1\,\ensuremath{v\sin i}$. Below this value, lines of both components merge.
%%_________________________________________________
\begin{figure*}[!htp]
        \centering
\resizebox{12cm}{!}{\includegraphics{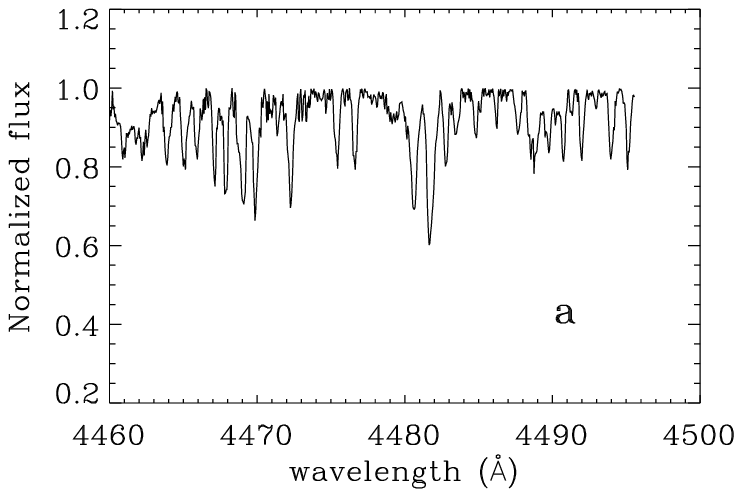}\includegraphics{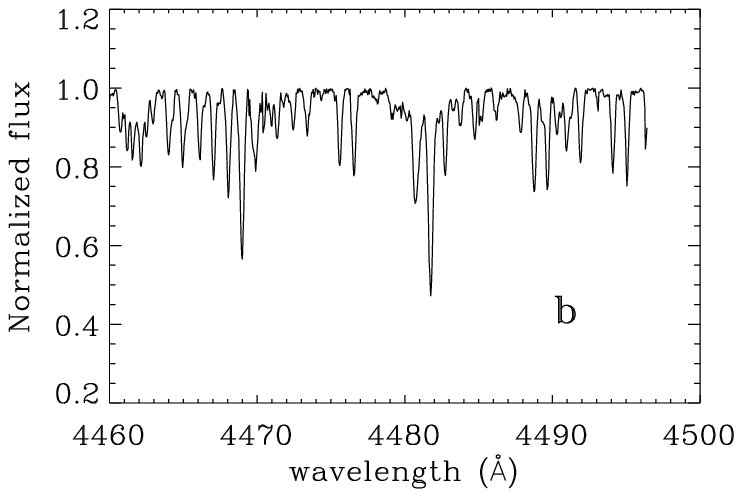}}  
\hfill
\parbox[b]{55mm}{  
\caption{The part of the spectrum of HD 10167, centered around
  \ion{Mg}{ii} 4481 (4460--4500\,\AA) is displayed for the two
  observed spectra of the star. Both panels present the
  binarity. Relative radial velocities are high enough compared to
  rotational broadening to allow to measure \ensuremath{v\sin i}\ for both
  components. Observation {\bf b} occurs nearly two years after
  observation {\bf a}.}
\label{sb2_1}
}
\end{figure*}

\begin{figure*}[!htp]
        \centering
\resizebox{\hsize}{!}{\includegraphics{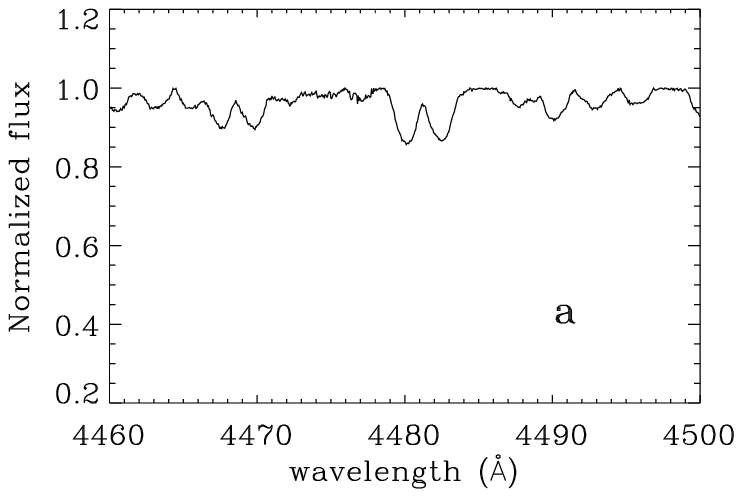}\includegraphics{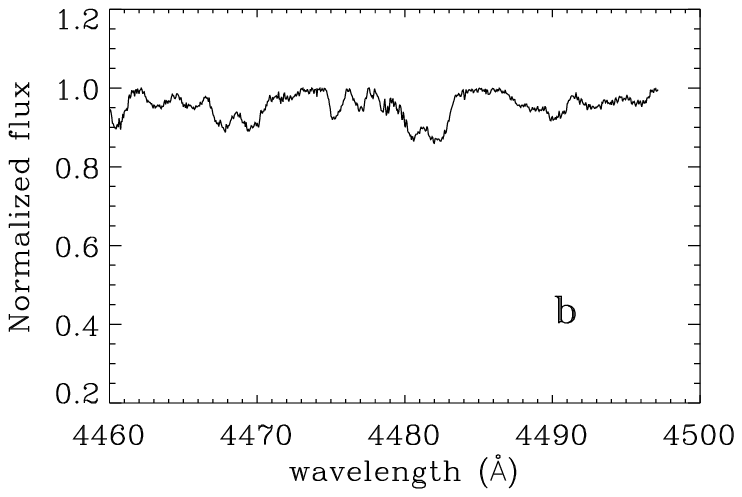}\includegraphics{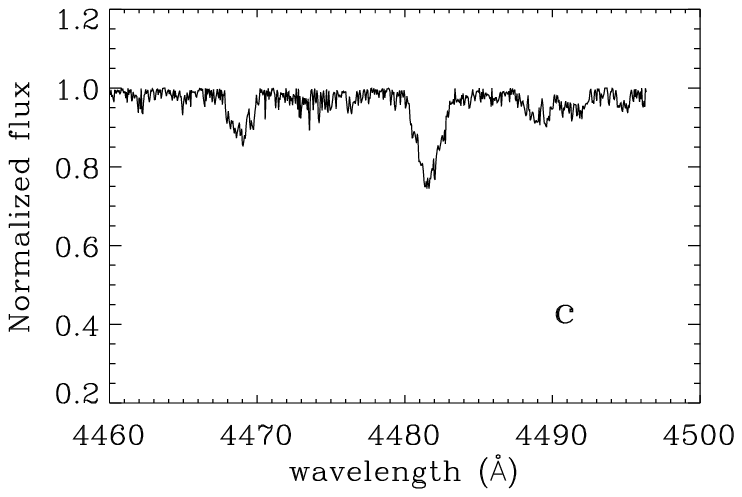}}  
        \caption{HD 18622 has been observed at three different times: {\bf a} HJD 2\,447\,790, {\bf b} 2\,448\,525 and {\bf c} 2\,448\,584. For each spectrum the region around \ion{Mg}{ii} 4481\,\AA\ is displayed. Relative radial velocities vary from about indiscernible components in panel {\bf c} to nearly 150\,\ensuremath{\mathrm{km}\,\mathrm{s}^{-1}}\ in {\bf a}. Relatively high rotational broadening makes the measurement of \ensuremath{v\sin i}\ difficult because of the ratio $\Delta V_\mathrm{r}\over\ensuremath{v\sin i}$, and derived rotational velocities are uncertain.}
        \label{sb2_2}
\end{figure*}

\begin{figure*}[!htp]
        \centering
\resizebox{12cm}{!}{\includegraphics{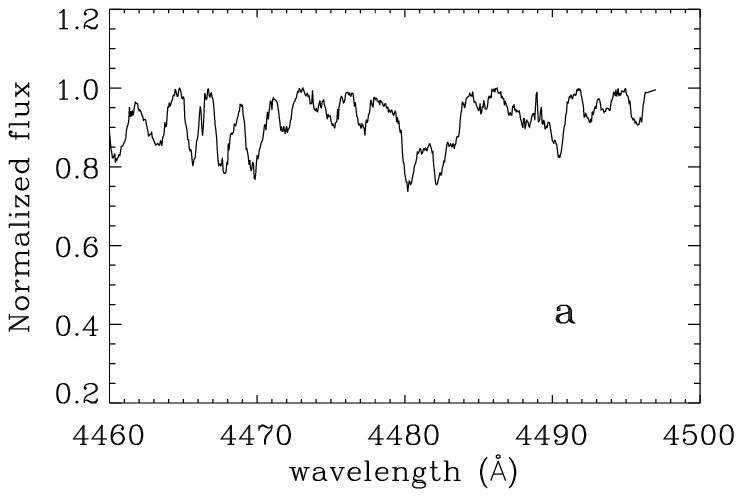}\includegraphics{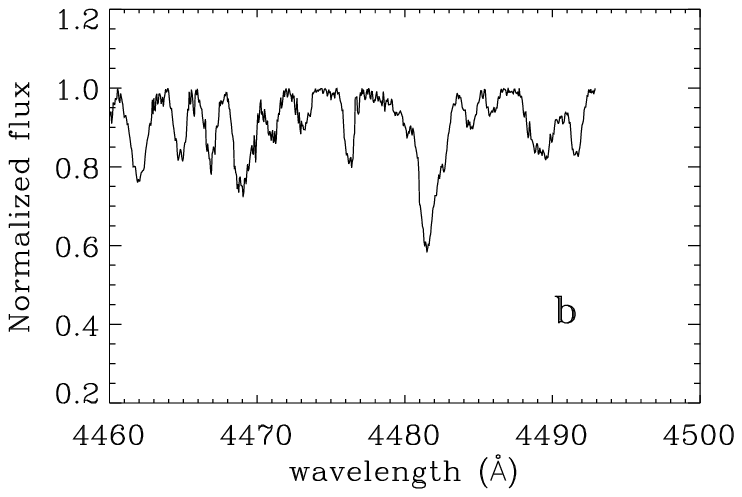}}  
\hfill
\parbox[b]{55mm}{  
\caption{HD 27346 spectra have been collected at two different orbital phases separated in time by 981 days. \ion{Mg}{ii} line shows clearly the two components in panel {\bf a}, whereas they are merged in {\bf b}.}
\label{sb2_3}
}
\end{figure*}

\begin{figure*}[!htp]
        \centering
\resizebox{\hsize}{!}{\includegraphics{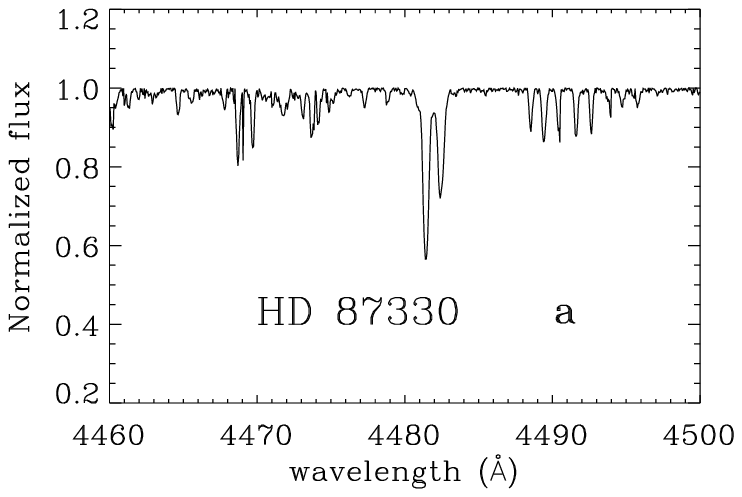}\includegraphics{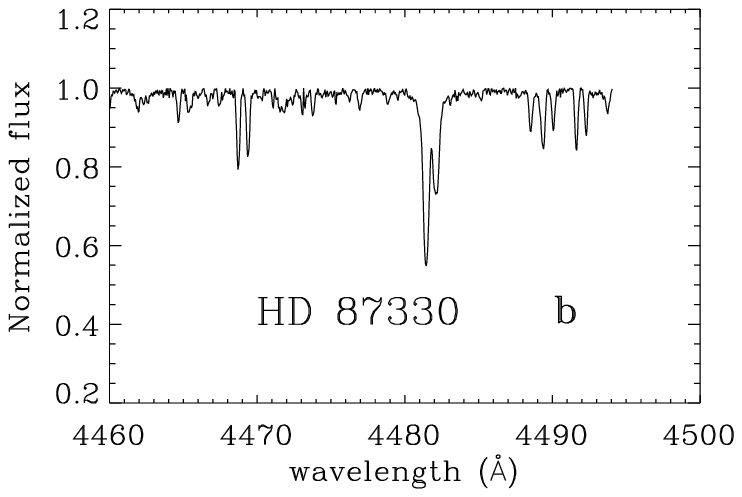}\includegraphics{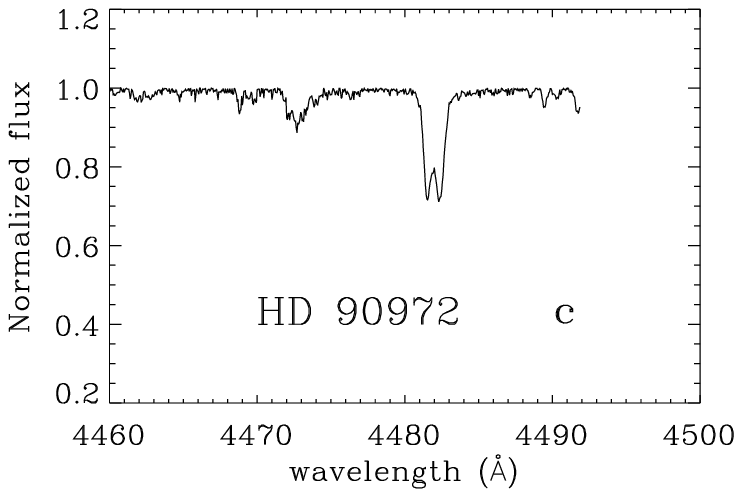}}  
\caption{{\bf a} and {\bf b} Observations of HD 87330 around \ion{Mg}{ii}, separated by almost three years. Low rotational broadening allows the measurement of \ensuremath{v\sin i}\ using weak metallic lines, whereas \ion{Mg}{ii} line of both components overlap, due to the intrinsic width of the doublet. {\bf c} Spectrum of the late B star HD 90972, with few metallic lines. The low difference of radial velocities makes the measurement of \ensuremath{v\sin i}\ uncertain.}
\label{sb2_4}
\end{figure*}

%%___________________________________________________

Table~\ref{SB2} displays the results for the stars in our sample which
exhibit an SB2 nature. We focus only on stars in which the spectral lines
of both component are separated. Spectral lines are identified by comparing the SB2 spectrum with a single star spectrum. Projected rotational velocities are given for each component when measurable, as well as  the difference in radial velocity $\Delta V_\mathrm{r}$ computed from the velocities given by \citet{Grr_99a}.
%%______________________________________________________________
\begin{table}[!htp]
\centering
\caption{Results of the \ensuremath{v\sin i}\ measurements for individual spectra of SB2 systems. When available, the \ensuremath{v\sin i}\ measurements are given for each component (A for the bluest and B for the reddest). The difference of radial velocity is also given, derived from Grenier et al. Dash (--) indicates a non-determined value: no double-peak CCF for $\Delta V_\mathrm{r}$ measurement, and $1.1\,\ensuremath{v\sin i}\lesssim\Delta V_\mathrm{r}\lesssim 3.4\,\ensuremath{v\sin i}$ for individual \ensuremath{v\sin i}\ measurement. When SB2 signature is not detectable a single \ensuremath{v\sin i}\ of merged lines is measured. Last column refers to the corresponding figures.}
\label{SB2}
\setlength\tabcolsep{5pt}
\begin{tabular}{rrlrrrc}
\hline
\multicolumn{1}{c}{HD}  & \multicolumn{1}{c}{HIP} & Spect. type & \multicolumn{2}{c}{\ensuremath{v\sin i}} & $\Delta V_\mathrm{r}$ & Fig.  \\
    &    &    & \multicolumn{2}{c}{(\ensuremath{\mathrm{km}\,\mathrm{s}^{-1}})}&\multicolumn{1}{c}{(\ensuremath{\mathrm{km}\,\mathrm{s}^{-1}})}  \\
    &    &    & \multicolumn{1}{c}{A}&\multicolumn{1}{c}{B}  \\
\hline
10167  &  7649  & F0V      &   17 &   14            &   80   & \ref{sb2_1}a.     \\ % ach0150
       &        &          &   11 &   13            &   62   & \ref{sb2_1}b.     \\ % far0046
18622  & 13847  & A4III+...&   71:&   74:           &  154   & \ref{sb2_2}a.     \\ % g24
       &        &          &  --  &   --            &  109   & \ref{sb2_2}b.     \\ % ohp0405
       &        &          &\multicolumn{2}{c}{83}  &   --   & \ref{sb2_2}c.     \\ % jac0060
27346  & 19704  & A9IV     &   35 &   35            &  135   & \ref{sb2_3}a.     \\ %    ap71
       &        &          &\multicolumn{2}{c}{36:} &   --   & \ref{sb2_3}b.     \\ %  ab07m 
87330  & 49319  & B9III/IV &   11 &    9            &   67   & \ref{sb2_4}a.     \\ %   bb98c
       &        &          &   10 &   10            &   45   & \ref{sb2_4}b.     \\ %     t54
90972  & 51376  & B9/B9.5V &   23:&   29:           &   54   & \ref{sb2_4}c.     \\ %     j72

\hline
\end{tabular}
\end{table}
%%__________________________________________________________

\begin{itemize}
\item \object{HD 10167} has given no indication of a possible duplicity up to now in the literature and is used as a photometric standard star \citep{Cou74}.
\item \object{HD 18622} ($\theta^1$~Eri) is a binary system for which HIPPARCOS measured the angular separation $\rho=8\farcs31{\scriptstyle\pm 0.003}$ and the difference of magnitude $\Delta Hp=1.09{\scriptstyle\pm 0.01}$ of the visual system, while our data concern the SB2 nature of the primary. 
\item \object{HD 27346} is suspected to be an astrometric binary
  on the basis of HIPPARCOS observations.
\item \object{HD 87330} was detected as a variable star by HIPPARCOS and its variability is possibly due to duplicity.
\item \object{HD 90972} ($\delta$~Ant) is a visual double system
  \citep{Pai_92} whose primary is an SB2 for which the \ensuremath{v\sin i}\ of both
  components are given in Table~\ref{SB2}. Grenier et al. do not
  identify $\delta$~Ant as an SB2 system but point to it as a certain double star on the basis of the CCF. It is worth noticing that their CCF is equivalent to a convolution with observed and synthetic spectra and the resulting profile is smoothed by both.
\end{itemize}

The magnesium doublet is perfectly suited to distinguish a spectral duplicity, so that spectral domain around 4481~\AA\ is displayed for SB2 systems is Figs.~\ref{sb2_1}, \ref{sb2_2}, \ref{sb2_3}, \ref{sb2_4}. However, the intrinsic width of the doublet increases its blend due to multiplicity whereas fainter lines can be clearly separated, and \ion{Mg}{ii} line is not used to derive \ensuremath{v\sin i}\ for SB2 systems.

Less obvious SB2 lie in our sample, but individually analyzing line
profiles one-by-one is not an appropriate method for detecting
them. Results about binarity for these spectra are however indicated in Grenier et al.

\subsubsection{Metallic shell stars}

\begin{figure*}[!htp]
        \centering
\resizebox{12cm}{!}{\includegraphics{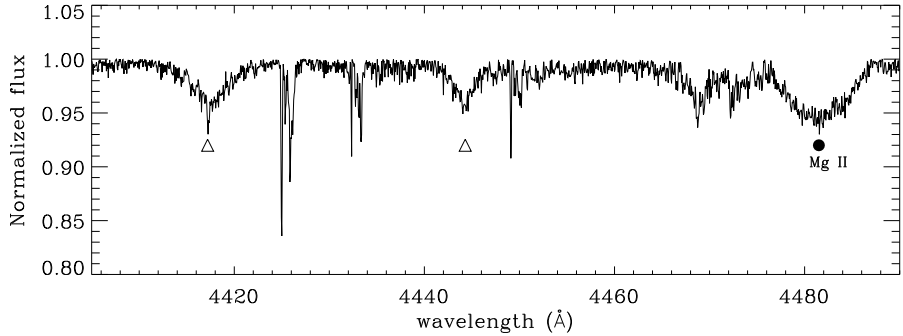}}  
\hfill
\parbox[b]{55mm}{  
\caption{Part of the spectrum of HD~225200 showing the rotationally broadened line \ion{Mg}{ii} 4481 (filled circle) and metallic lines exhibiting the signature of the shell as sharp core and extended wings (open triangle).}
\label{shell_1}
}
\end{figure*}

The specific ``shell'' feature in stars with a circumstellar envelope
is characterized by double emission and central absorption in hydrogen
lines. This characteristic is likely a perspective effect, as
suggested by \citet{Slk79}, and shell-type lines occur at high
inclination $i$ when line of sight intersects with the disk-like
envelope. For our purpose, \ensuremath{v\sin i}\ determination, critical effect is
due to metallic shell stars, where shell-type absorption not only
occurs in Balmer series but also in metallic lines. Our candidate
lines exhibit a broad profile, indicating rapid rotation of the
central star, a high inclination of the line of sight, and a
superimposed sharp absorption profile originating in the circumstellar
envelope (Fig.~\ref{shell_1}). Metallic shell-type lines arise when
perspective effect is more marked than for hydrogen shell stars
\citep{Brt86}. Measurement of \ensuremath{v\sin i}\ requires a line profile from the
central star photosphere only, and not polluted by absorption caused by the circumstellar envelope which does not reflect the rotation motion.
%%_________________________________________________
%%_________________________________________________

Derived \ensuremath{v\sin i}\ for the metallic shell stars present in our sample are listed in Table~\ref{shell}. These stars are already known as shell stars. HD~15004 (71~Cet) and HD~225200 are further detailed by Gerbaldi et al. In our spectral range, magnesium multiplet \ion{Mg}{ii} 4481 is the only measurable line. 
%%______________________________________________________________
\begin{table}[!htp]
\centering
\caption{Results of the \ensuremath{v\sin i}\ measurements for individual spectra of metallic shell stars.}
\label{shell}
\begin{tabular}{rrlc@{\hspace*{1mm}}c@{\hspace*{1mm}}}
\hline
\multicolumn{1}{c}{HD}  & \multicolumn{1}{c}{HIP} & Spectral type & \ensuremath{v\sin i}\\
                        &                         &               & (\ensuremath{\mathrm{km}\,\mathrm{s}^{-1}})\\
\hline
 15004 & 11261 & A0III     & 249 \\ 
 24863 & 18275 & A4V       & 249 \\                 
 38090 & 26865 & A2/A3V    & 204 \\ 
 88195 & 49812 & A1V       & 236 \\ 
 99022 & 55581 & A4:p      & 236 \\                  
       &       &           & 236 \\                     
       &       &           & 249 \\                
225200 &   345 & A1V       & 345 \\     

\hline
\end{tabular}
\end{table}
%%__________________________________________________________

%==========================================
\subsection{Comparison with existing data}
%==========================================
\label{comparison}

%%_________________________________________________
\begin{figure*}[!htp]
\resizebox{12cm}{!}{\includegraphics{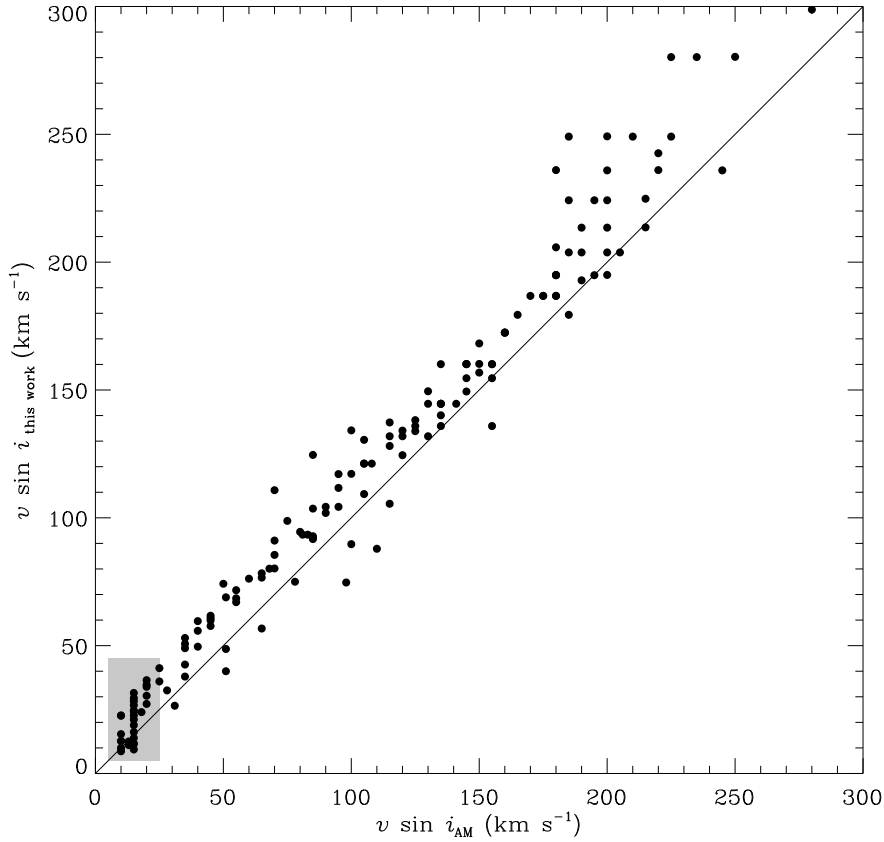}}
\hfill
\parbox[b]{55mm}{  
\caption{Comparison between \ensuremath{v\sin i}\ values from this work and from
  \citet[][ AM]{AbtMol95} for the 160 common stars. The solid line
  stands for the one-to-one relation. The grey box encompasses the points of low \ensuremath{v\sin i}, for which the relation has a much higher local  slope and produces an overestimation of the global slope.}
\label{comp-vsini1}
}
\end{figure*}
%%___________________________________________________

%%_________________________________________________
\begin{figure*}[!htp]
\resizebox{12cm}{!}{\includegraphics{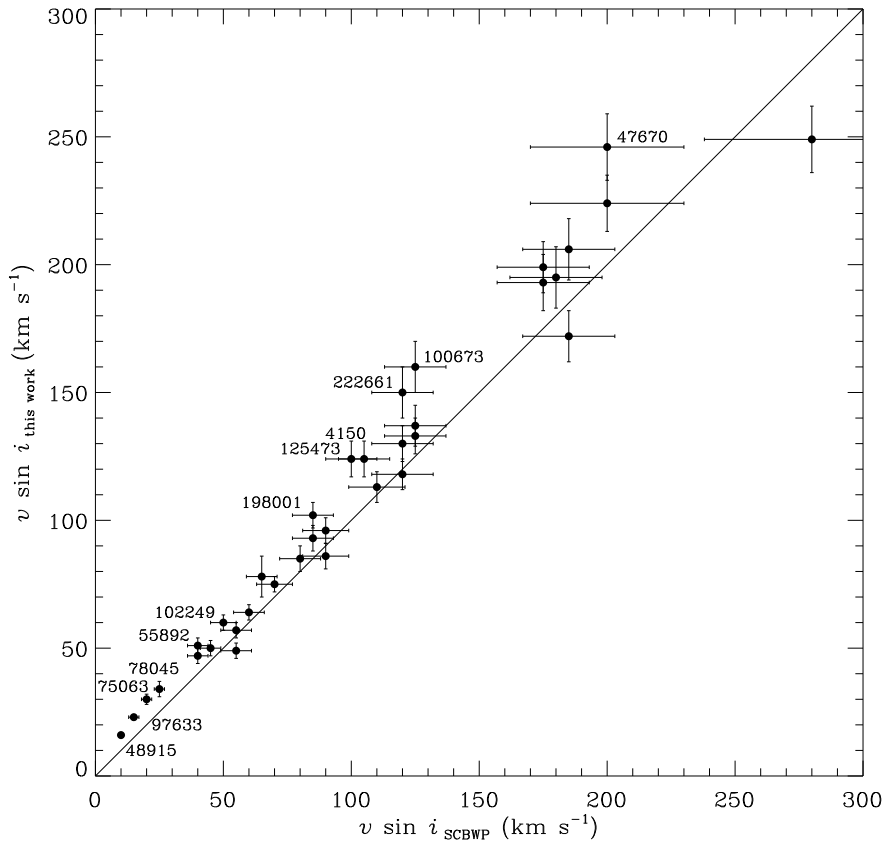}}
\hfill
\parbox[b]{55mm}{  
\caption{Comparison between \ensuremath{v\sin i}\ data from this work and from
  Slettebak et al. The solid line stands for the one-to-one
  relation. The 35 standard stars are plotted with error bar on both
  axes (see text). The stars that deviate most from the one-to-one
  relation have their HD number indicated and are summarized in Table~\ref{vsini_litt}.}
\label{comp-vsini}
}
\end{figure*}
%%___________________________________________________

%[SLETTEBAK et al.]
 
The most homogeneous large data set of rotational velocities for
A-type stars which has been provided up to now is that of \citet{AbtMol95},
who measured \ensuremath{v\sin i}\ for about 1700 A-type stars in the northern
hemisphere. 
The intersection with our southern sample includes 160 stars. The
comparison of the \ensuremath{v\sin i}\ (Fig.~\ref{comp-vsini1}) shows that our determination is higher on average than
the velocities derived by Abt \& Morrell (AM). The linear relation
given by GaussFit is:
\begin{equation}
\label{Ram-AM}
\ensuremath{v\sin i}_\mathrm{this\;work} = 1.15{\scriptstyle\pm 0.03}\,\ensuremath{v\sin i}_\mathrm{AM}+2.1{\scriptstyle\pm 0.8}.
\end{equation}

Abt \& Morrell based their measurements on the scale established by SCBWP, who built a new calibration FWHM-\ensuremath{v\sin i}, replacing the old system and leading to values 5\,\% smaller on average for A-F stars.

There are 35 stars in common between our sample and the standard stars of SCBWP. It is worth emphasizing that among these 35 stars, only one third has a gaussian CCF in the study of Grenier et al. Moreover there is an SB2 system (HD~18622) and almost one half of this group is composed of suspected or probable multiple stars, on the basis of their CCF.

Figure~\ref{comp-vsini} displays the \ensuremath{v\sin i}\ derived in this paper versus the \ensuremath{v\sin i}\ from SCBWP for the 35 standard stars in common. The solid line represents the one-to-one relation. A clear trend is observed: \ensuremath{v\sin i}\ from SCBWP are on average 10 to 12\,\% lower. A linear least squares fit carried out with GaussFit on these values makes the systematic effect explicit:

\begin{equation}
\label{Ram-Slk}
\ensuremath{v\sin i}_\mathrm{this\;work} = 1.04{\scriptstyle\pm 0.02}\,\ensuremath{v\sin i}_\mathrm{SCBWP}+6.1{\scriptstyle\pm 1.0}.
\end{equation}

The relation is computed taking into account the error bars of both sources. The error bars on the values of SCBWP are assigned according to the accuracy given in their paper (10\,\% for $\ensuremath{v\sin i}<200\,\ensuremath{\mathrm{km}\,\mathrm{s}^{-1}}$ and 15\% for $\ensuremath{v\sin i}\geq 200\,\ensuremath{\mathrm{km}\,\mathrm{s}^{-1}}$). Our error bars are derived from the formal error found in section~\ref{precision} (Eq.~\ref{sigma3}).

The difference between the two relations, (Eq.~\ref{Ram-AM}) and
(Eq.~\ref{Ram-Slk}), concerns mainly the low \ensuremath{v\sin i}\ region. When low \ensuremath{v\sin i}\ from Abt
\& Morrell, $<25$~\ensuremath{\mathrm{km}\,\mathrm{s}^{-1}}, are not taken into account (grey box in
Fig.~\ref{comp-vsini1}), the relation given by GaussFit between
\ensuremath{v\sin i}\ from Abt \& Morrell and this work becomes:
\begin{equation}
\label{Ram-AM2}
\ensuremath{v\sin i}_\mathrm{this\;work} = 1.08{\scriptstyle\pm 0.02}\,\ensuremath{v\sin i}_\mathrm{AM}+5.3{\scriptstyle\pm 1.8},
\end{equation}which is almost identical to the relation with SCBWP data (Eq.~\ref{Ram-Slk}).
 
\begin{table*}[!htp]
\begin{center}
\caption{Highlight of the discrepancy between \ensuremath{v\sin i}\ values from SCBWP and ours (standard deviation of our measurement is indicated; dash ``--'' stands for only one measurement). Comparison with data from the literature for the twelve stars that exhibit the largest differences. \ensuremath{v\sin i}\ are classified in three subgroups according to the way they are derived: by-product of a spectrum synthesis, frequency analysis of the lines profiles or infered from a FWHM-\ensuremath{v\sin i}\ relation independent from SCBWP's one. Flags from HIPPARCOS catalogue are indicated: variability flag H52 (C: constant, D: duplicity-induced variability, M: possibly micro-variable, U: unsolved variable, --: no certain classification) and double annex flag H59 (O: orbital solution, G: acceleration terms, --: no entry in the Double and Multiple Systems Annex). The shape of  the CCF found by Grenier et al. is also given (0: symmetric and gaussian peak, 1: SB2, 2: certain double, near spectral types, 3: certain double, A-B type with faint F-G component, 4: probable double, 5: suspected double, 6: probable multiple system, 7: certain shell star, 8: suspected shell star, 9: wide and irregular peak, 10: wide peak of B star (few lines)). }
\label{vsini_litt}
\begin{tabular}{lrlrr@{\hspace*{1mm}}llllccc}
\hline
\multicolumn{1}{c}{Name} & \multicolumn{1}{c}{HD} &
\multicolumn{1}{c}{Sp. type} &  \multicolumn{6}{c}{\ensuremath{v\sin i}\ (\ensuremath{\mathrm{km}\,\mathrm{s}^{-1}})}
& \multicolumn{2}{c}{{\sc hipparcos}} & CFF             \cr
   &    &      &  {\sc scbwp} & \multicolumn{2}{c}{this
     work}&\multicolumn{3}{c}{\hbox{\raisebox{0.3em}{\vrule depth 0pt
         height 0.4pt width 2.5cm} literature \raisebox{0.3em}{\vrule
         depth 0pt height 0.4pt width 2.5cm}}} & H52 & H59\cr
   &     &      &    &    &      & \multicolumn{1}{c}{spec. synth.}& \multicolumn{1}{c}{freq. analysis}& \multicolumn{1}{c}{FWHM}\cr
\hline
$\eta$~Phe     &  4150& A0IV & $105$ &$124$& ${\scriptstyle\pm 0}$  & -- & -- & -- & C & G & 4\\
$\nu$~Pup      & 47670& B8III &  $200$ &$246$& ${\scriptstyle\pm 7}$ & -- &
-- & --  & U & -- & 5\\
$\alpha$~CMa   & 48915& A0m...  &  $10$ & $16$& ${\scriptstyle\pm 1}$ & $16{\scriptstyle\pm 1}^{(1)}\;16^{(2)}$ & $17^{(4)}\;16.9^{(5)}$ & -- & -- & -- & 0\\
               &      &   &       &     &   & $16.2^{(3)}$ &  $19^{(6)}\;15.3{\scriptstyle\pm 0.3}^{(7)}$ \\
$QW$~Pup       & 55892& F0IV  &  $40$ & $51$& ${\scriptstyle\pm 8}$ &
--     &       --      & $50^{(8)}$ & M & -- & 4\\
$a$~Vel        & 75063& A1III  &  $20$ & $30$& ${\scriptstyle\pm 2}$ & -- & -- & -- & -- & -- & 0\\
$\alpha$~Vol   & 78045& Am  &  $25$ & $34$& ${\scriptstyle\pm 2}$ & $45^{(9)}$ &       --      &    -- & C & -- & 0\\
$\theta$~Leo   & 97633& A2V  &  $15$ & $23$& ${\scriptstyle\pm 1}$ & $21^{(2)}\;22.1^{(3)}$ & -- & $23^{(10)}$ & -- & -- & 0\\
$A$~Cen        &100673& B9V &  $125$ &$160$& -- & -- & -- & -- & C & -- & 10\\ 
$\lambda$~Mus  &102249& A7III & $50$ & $60$& ${\scriptstyle\pm 2}$ &
-- & -- & $60^{(11)}$ & C & O & 0\\
$\psi$~Cen     &125473& A0IV &  $100$ &$124$& ${\scriptstyle\pm 2}$ & $132^{(9)}$&       --      &-- & -- & -- & 5\\
$\epsilon$~Aqr &198001& A1V  &  $85$ &$102$& -- & $108.1^{(3)}\;95^{(12)}$& -- &  -- & -- & -- & 0\\
$\omega^2$~Aqr &222661& B9V &  $120$ &$150$& -- & -- & -- & -- & C & -- & 4\\ 
\hline
\end{tabular}
\end{center}
\begin{tabular}{llll}
$^{(1)}$ \citet{Kuz_77}  & $^{(5)}$ \citet{Deg77}  & $^{(9)}$  \citet{Hor_99}\\
$^{(2)}$ \citet{Lee89}   & $^{(6)}$ \citet{Raa_89} & $^{(10)}$ \citet{Fel98}\\
$^{(3)}$ \citet{Hil95}   & $^{(7)}$ \citet{Drs_90} & $^{(11)}$ \citet{Noi_84}\\
$^{(4)}$ \citet{Smh76}   & $^{(8)}$ \citet{Ban90}  & $^{(12)}$ \citet{Dun_97}\\
\end{tabular}
\end{table*}

For slow rotational velocities, the discrepancy far exceeds the
estimate of observational errors. Figure~\ref{comp-vsini} also shows
the stars which deviate the most from the one-to-one relation. These
twelve stars, for which the error box around the point does not
intersect with the one-to-one relation, are listed in
Table~\ref{vsini_litt} with different rotational velocity
determinations gathered from the literature.
 Their large differences together with comparison to other data
allow us to settle on which source carries the systematic effect. 
Without exception, all data gathered from the literature and
listed in Table~\ref{vsini_litt} are systematically higher than the
corresponding SCBWP's \ensuremath{v\sin i}\ and for the majority of the listed stars, data from the literature are consistent with our \ensuremath{v\sin i}\ determinations.
These stars are further detailed in the Appendix~\ref{Notes3}.

%%%%%%%%%%%%%%%%%%%%%%%%%%%%%%%%
%%%%%%%%(  DISCUSSION  )%%%%%%%%
%%%%%%%%%%%%%%%%%%%%%%%%%%%%%%%%
\section{Discussion and conclusion}

The selection of several suitable spectral lines and the evaluation of 
their reliability as a function of broadening and effective
temperature allows the computation of \ensuremath{v\sin i}\ over the whole spectral range of 
A-type stars and a robust estimate of the associated relative error.

Up to 150~\ensuremath{\mathrm{km}\,\mathrm{s}^{-1}}, a statistical analysis indicates that the standard
deviation is about 6~\% of the \ensuremath{v\sin i}.
It can be seen, in both Figs.~\ref{comp-vsini1} and \ref{comp-vsini},
that the dispersion increases beyond 180~\ensuremath{\mathrm{km}\,\mathrm{s}^{-1}} approximately, when
comparing rotational velocities to previous determination by Abt \&
Morrell and SCBWP. SCBWP estimate a larger uncertainty for rotational velocities higher than
200~\ensuremath{\mathrm{km}\,\mathrm{s}^{-1}}; nevertheless our precision estimation for a 200~\ensuremath{\mathrm{km}\,\mathrm{s}^{-1}}
\ensuremath{v\sin i}\ is extrapolated from Fig.~\ref{sigma-vsini1}. Errors
may thus be larger, due to the sampling in Fourier space, which is proportional to
$(\ensuremath{v\sin i})^2$. 

In addition, determination of continuum level induces a
systematic underestimation of \ensuremath{v\sin i}\ that reaches about 5 to 10~\%
depending on the lines and broadening.

Gravity darkening \citep[von Zeipel effect,][]{VZl25} is not taken into account in
this work. \citet{HapStr68} quantify this effect, showing that \ensuremath{v\sin i}\
could be 15 to 40~\% too small if gravity darkening is neglected for
stars near break-up velocity. Nevertheless, in a recent work \citep{Shn00}, this
effect is revised downwards and found to remain very small as long as
angular velocity is not close to critical velocity ($\omega < 0.8$): it induces an
underestimation
lower than 1~\% of the FWHM. 
In our observed sample, 15 stars (with spectral type from B8V to
A1V) have $\ensuremath{v\sin i}>250$~\ensuremath{\mathrm{km}\,\mathrm{s}^{-1}}.
According to their radii and masses, derived from empirical
calibrations \citep{HasHee81}, their critical velocities $v_\mathrm{c}$ are higher
than 405~\ensuremath{\mathrm{km}\,\mathrm{s}^{-1}}\ (Zorec, private communication).
Only seven stars have a high \ensuremath{v\sin i}, so that $\ensuremath{v\sin i}/v_\mathrm{c}> 0.7$. 
The fraction of stars rotating near their break-up velocity remains
very small, probably lower than 2~\% of the sample size.

\paragraph{}A systematic shift is found with the values from the
catalogue of \citet{AbtMol95}. This difference arises from the use of the
calibration relation from SCBWP, for which a similar shift is
found. The discrepancy observed with standard \ensuremath{v\sin i}\ values given by
SCBWP has already been mentioned in the literature.
\citet{Raa_89} point out a similar shift with respect to the \ensuremath{v\sin i}\ from SCBWP. They suppose that the discrepancy could come from the models SCBWP used to compute theoretical FWHM of the \ion{Mg}{ii} line.
\citet{BrnVen97} derived \ensuremath{v\sin i}\ for early-type stars. For low \ensuremath{v\sin i}\ (up to $\sim 60$\,\ensuremath{\mathrm{km}\,\mathrm{s}^{-1}}), their values are systematically higher than those of SCBWP. They attribute this effect to the use of the models from \citet{CosSon77} by SCBWP; they assert that using the modern models of \citet{Cos_91}, to derive \ensuremath{v\sin i}\ from FWHM, eliminates the discrepancy. 
Fekel (private communication) also finds this systematic effect between values from \citet{AbtMol95}, which are directly derived from the SCBWP's calibration, and the \ensuremath{v\sin i}\ he measured using his own calibration \citep{Fel97}.

In addition, some stars used as \ensuremath{v\sin i}\ standards turn out to be multiple
systems or to have spectral features such that their status as a
standard is no longer valid. The presence of these ``faulty'' objects in the standard star sample may introduce biases in the \ensuremath{v\sin i}\ scale. There is no doubt that the list of standards established by SCBWP has to be revised.

The above comparisons and remarks lead us to call into question the
\ensuremath{v\sin i}\ values of the standard stars from SCBWP.

This paper is a first step, and a second part will complete these data 
with a northern sample of A-type stars.

%%%%%%%%%%%%%%%%%%%%%%%%%%%%%%%%
%%%%%%%%(    THANKS    )%%%%%%%%
%%%%%%%%%%%%%%%%%%%%%%%%%%%%%%%%

\begin{acknowledgements}
We are very grateful to Dr~M.~Ramella for providing us the computer
program used to derive the \ensuremath{v\sin i}. We also thank the
referee, Prof.~J.~R.~de~Medeiros, for his several helpful suggestions. Precious advice on statistical
analysis was kindly given by Dr~F.~Arenou and was of great utility. 
We want to acknowledge Dr~F.~C.~Fekel for his help in comparing $v\sin i$
with data from the literature.
Finally, we are thankful to B.~Tilton for her careful reading of the
manuscript. This work made use of the SIMBAD database, operated at CDS, Strasbourg, France.
\end{acknowledgements}

%%%%%%%%%%%%%%%%%%%%%%%%%%%%%%%%
%%%%%%%%(  APPENDIX    )%%%%%%%%
%%%%%%%%%%%%%%%%%%%%%%%%%%%%%%%%
\appendix
\section{notes on stars with uncertain rotational velocity}
\label{Notes0}
\subsection{Stars with no selected line} 
For a few spectra, all the measurable lines were discarded either a
priori from Table~\ref{skewness} or a posteriori because of a
distorted FT. These stars appear in Table~\ref{results} with an
uncertain \ensuremath{v\sin i}\ (indicated by a colon) and a flag ``NO''. They are detailed below:

\begin{itemize}
\item  HD~1978 is rapid rotator whose \ion{Mg}{ii} line FT profile is
  distorted ($\ensuremath{v\sin i} = 110:$~\ensuremath{\mathrm{km}\,\mathrm{s}^{-1}}).

\item  HD~41759, HD~93905 and HD~99922 have a truncated spectrum: only
  two thirds of the spectral range are covered (from 4200 to
  4400\,\AA). Half of the selected lines are thus unavailable and
  estimation of the \ensuremath{v\sin i}\ is then given, as an indication, by lines
  that would have normally been discarded by the high skewness of
  their synthetic profile. ($\ensuremath{v\sin i} = 215:,\; 85:,\; 65:$~\ensuremath{\mathrm{km}\,\mathrm{s}^{-1}}\ respectively)

\item  HD~84121 is a sharp-lined A3IV star ($\ensuremath{v\sin i} = 10:$~\ensuremath{\mathrm{km}\,\mathrm{s}^{-1}}). It is resolved by
  HIPPARCOS as a binary system (separation $0\farcs 125\,\pm 0\farcs 006$ and magnitude
  difference $\Delta Hp = 0.61 \pm 0.31$~mag) and highly
  suspected to hide a third component \citep{Som99}.

\item  HD~103516 is a supergiant A3Ib star ($\ensuremath{v\sin i} = 20:$~\ensuremath{\mathrm{km}\,\mathrm{s}^{-1}}).

\item  HD~111786 is detected binary by
  \citet{Faa_97} ($\ensuremath{v\sin i} = 45:$~\ensuremath{\mathrm{km}\,\mathrm{s}^{-1}}).

\item  HD~118349 is a A7-type star indicated as variable in photometry, due to
  duplicity, by HIPPARCOS ($\ensuremath{v\sin i} = 100:$~\ensuremath{\mathrm{km}\,\mathrm{s}^{-1}}).

\end{itemize}

\subsection{Stars with high external error}
The following stars have a mean \ensuremath{v\sin i}\ whose associated standard deviation is higher than 15\,\% of the \ensuremath{v\sin i}. This dispersion may either come from differences from one spectrum to another or lie in a single spectrum.

Some of the stars whose spectrum has been collected several times show
different \ensuremath{v\sin i}\ from one spectrum to another (flag ``SS'' in
Table~\ref{results}). These differences could be related to intrinsic
variations in the spectrum itself. Other stars present a high
dispersion in the measures from lines in a single spectrum (flag ``LL'' in Table~\ref{results}). These stars are detailed in appendices \ref{Notes1} and \ref{Notes2} respectively.

\subsubsection{Variations from spectrum to spectrum}
\label{Notes1}
\begin{itemize}
\item HD~55892 has two very different \ensuremath{v\sin i}\ derived from its two
  collected spectra : 45 and 56\,\ensuremath{\mathrm{km}\,\mathrm{s}^{-1}}. Probable double in \citet{Grr_99a}.
\item HD~74461 $73\,\ensuremath{\mathrm{km}\,\mathrm{s}^{-1}}\pm 15\,\ensuremath{\mathrm{km}\,\mathrm{s}^{-1}}$ and $93\,\ensuremath{\mathrm{km}\,\mathrm{s}^{-1}}\pm 6\,\ensuremath{\mathrm{km}\,\mathrm{s}^{-1}}$. Probable double in \citet{Grr_99a}.
\item HD~87768 shows two different values in the two spectra: 90 and
  112\,\ensuremath{\mathrm{km}\,\mathrm{s}^{-1}}. It is indicated as a certain double star by \citet{Grr_99a}, whose primary is an A star and secondary
  should be a faint F-G star. \citet{AbtWih94}, point to
  it as an SB2 system on the basis of the study of its radial velocity.
\item HD~174005 has two very different \ensuremath{v\sin i}\ derived from its two
  collected spectra ($129\pm 5$ and $66\pm 7$\,\ensuremath{\mathrm{km}\,\mathrm{s}^{-1}}). It is classified as a certain
  double star, with components of similar spectral types, by \citet{Grr_99a}.
\item HD~212852 shows two different values in the two spectra: 93 and
  121\,\ensuremath{\mathrm{km}\,\mathrm{s}^{-1}}. It is suspected double by \citet{Grr_99a}.
\end{itemize}

\subsubsection{Variations from line to line}
\label{Notes2}
\begin{itemize}
\item HD~40446 is a A1Vs star whose \ensuremath{v\sin i}\ is found as $27\pm 5$~\ensuremath{\mathrm{km}\,\mathrm{s}^{-1}}. The
  stochastic motion solution in HIPPARCOS
  data may suggest a possible multiplicity. 
\item HD~109074 is a A3V star with $\ensuremath{v\sin i} = 84\pm 15$~\ensuremath{\mathrm{km}\,\mathrm{s}^{-1}}. HIPPARCOS astrometric
  solution comprises acceleration terms, which could indicate a multiplicity.
\end{itemize}

\section{notes on {\boldmath \ensuremath{v\sin i}} standard stars with discrepant rotational velocity}
\label{Notes3}
Standard stars from SCBWP that are listed in Table~\ref{vsini_litt},
are now detailed:

\begin{itemize}

\item \object{$\eta$~Phe} (HD~4150) is among the A0 dwarf stars
  investigated by \citet{Gei_99}. They suspect
  $\eta$~Phe to be a binary system on the basis of the fit between the
  observed and the computed spectrum, as do Grenier et al. on the
  basis of the CCF of the spectrum.

\item \object{$\nu$~Pup} (HD~47670) is a late B giant star. It is part
  of the sample studied by \citet{Bae89a,Bae89b} who searches for line
  profile variability. He detects roughly central quasi-emission bumps
  in the rotationally broadened \ion{Mg}{ii} absorption line. This
  feature could be caused by the change in effective gravity from equator
  to poles and the associated temperature differences because of fast
  rotation. \citet{Ris_99} tone down this result but conclude
  nevertheless that there is strong evidence that this star is a not previously recognized bright Be star. 
 $\nu$~Pup was suspected to be a $\beta$ Cephei star
\citep{Shw75} and is newly-classified as an irregular variable on the basis of the HIPPARCOS photometric observations.

\item \object{Sirius} ($\alpha$~CMa, HD~48915) has a
  $\ensuremath{v\sin i}=10$\,\ensuremath{\mathrm{km}\,\mathrm{s}^{-1}}\ in the catalogue of SCBWP. Several authors have given larger values derived from approximatively the same spectral domain. \citet{Smh76} finds 17\,\ensuremath{\mathrm{km}\,\mathrm{s}^{-1}}\ and \citet{Kuz_77} 16\,\ensuremath{\mathrm{km}\,\mathrm{s}^{-1}}\ $\pm 1$\,\ensuremath{\mathrm{km}\,\mathrm{s}^{-1}}. \citet{Deg77}, using Bessel functions, finds 16.9\,\ensuremath{\mathrm{km}\,\mathrm{s}^{-1}}. \citet{Lee89}, analyzing abundance anomalies in A stars, derives the \ensuremath{v\sin i}\ of Sirius as an optimum fit over the spectral range: 16\,\ensuremath{\mathrm{km}\,\mathrm{s}^{-1}}. \citet{Drs_90} using Fourier techniques on high resolution and high signal-to-noise ratio  spectra give 15.3\,\ensuremath{\mathrm{km}\,\mathrm{s}^{-1}}\ $\pm 0.3$\,\ensuremath{\mathrm{km}\,\mathrm{s}^{-1}}. \citet{Hil95} studies a dozen A-type stars using spectral synthesis techniques to make an abundance analysis; fitting the spectra (four 65\,\AA\ wide spectral regions between 4500 and 5000\,\AA) he finds 16.2\,\ensuremath{\mathrm{km}\,\mathrm{s}^{-1}}\ as the \ensuremath{v\sin i}\ of Sirius.

\item \object{$QW$~Pup} (HD~55892) is an early F-type star. It belongs to the $\gamma$~Dor class of pulsating variable stars \citep{Kae_99}. \citet{Ban90}, studying lithium depletion, determines Li abundances and rotational velocities for a sample of nearly 200 F-type stars. Using her own FWHM--\ensuremath{v\sin i}\ calibration, she finds 50\,\ensuremath{\mathrm{km}\,\mathrm{s}^{-1}}\ for $QW$~Pup, which gets closer to our determination. HIPPARCOS results show that $QW$~Pup is a possible micro-variable star.

\item \object{$a$~Vel} (HD~75063) in an early A-type star. It is part of the sample of IRAS data studied by \citet{Ton_97} in the aim of detecting circumstellar dust shells. They do not rule out that $a$~Vel may have such a feature. No relevant \ensuremath{v\sin i}\ data have been found for this star.

\item \object{$\alpha$~Vol} (HD~78045) is a dusty A star on the basis of IRAS data \citep{Chg_92}. \citet{Hor_99} measure a \ensuremath{v\sin i}\ significantly larger than the \ensuremath{v\sin i}\ from SCBWP and even our determination.

\item \object{$\theta$~Leo} (HD~97633) is a candidate constant velocity A star. \citet{Fel98} monitors it and attributes to it a $\ensuremath{v\sin i}=23\,\ensuremath{\mathrm{km}\,\mathrm{s}^{-1}}$. \citet{Lee89} and \citet{Hil95} respectively measured it at 21 and 22.1\,\ensuremath{\mathrm{km}\,\mathrm{s}^{-1}}. These values are significantly higher than 15\,\ensuremath{\mathrm{km}\,\mathrm{s}^{-1}}\ found by SCBWP.

\item \object{$A$~Cen} (HD~100673) is a B-type emission line star.

\item \object{$\lambda$~Mus} (HD~102249) has been measured by \citet{Noi_84} who derive its \ensuremath{v\sin i}\ using the CCF and a calibration as described in \citet{Str_84}. 
In HIPPARCOS data, $\lambda$~Mus is a binary star for which orbital parameters are given: period $P = 453\,\mathrm{d} \pm 8$\,d, inclination $i=134\,\degr \pm 8$\,\degr, semi-major axis of photocentre orbit $a_0 =6.31\,\mathrm{mas} \pm 1.05$\,mas.

\item \object{$\psi$~Cen} (HD~125473) is a dusty A-star \citep{Chg_92}. The rotational velocity derived by \citet{Hor_99} agrees with our determination, 30\,\% larger than SCBWP.

\item \object{$\epsilon$~Aqr} (HD~198001) has a $\ensuremath{v\sin i}=85$\,\ensuremath{\mathrm{km}\,\mathrm{s}^{-1}}\ according to SCBWP, much smaller than the value in this work. The velocity derived by Hill is consistent with ours, taking into account the uncertainty of the measurements. \citet{Dun_97} found 95\,\ensuremath{\mathrm{km}\,\mathrm{s}^{-1}}\ by fitting the observed spectrum with a synthetic one.

\item \object{$\omega^2$~Aqr} (HD~222661) has, to our knowledge, no further determination of the \ensuremath{v\sin i}, independent of the SCBWP's calibration. 

\end{itemize}

%%%%%%%%%%%%%%%%%%%%%%%%%%%%%%%%%
%%%%%%%%| BIBLIOGRAPHY |%%%%%%%%%
%%%%%%%%%%%%%%%%%%%%%%%%%%%%%%%%%

\end{document}